\documentclass[10pt]{article}
\usepackage{ametsoc2col}

\newcommand{\myabstract}{Three dimensional (3D) Finite Time Lyapunov Exponents (FTLEs) are computed from numerical simulations of a freely evolving mixed layer (ML) front in a zonal channel undergoing baroclinic instability. The 3D FTLEs show a complex structure, with features that are less defined than the two-dimensional (2D) FTLEs, suggesting that stirring is not confined to the edges of vortices and along filaments and posing significant consequences on mixing. The magnitude of the FTLEs is observed to be strongly determined by the vertical shear. A scaling law relating the local FTLEs and the nonlocal density contrast used to initialize the ML front is derived assuming thermal wind balance. The scaling law only converges to the values found from the simulations within the pycnocline, while it displays differences within the ML, where the instabilities show a large ageostrophic component. The probability distribution functions of 2D and 3D FTLEs are found to be non Gaussian at all depths. In the ML, the FTLEs wavenumber spectra display -1 slopes, while in the pycnocline, the FTLEs wavenumber spectra display -2 slopes, corresponding to frontal dynamics. Close to the surface, the geodesic Lagrangian Coherent Structures (LCSs) reveal a complex stirring structure, with elliptic structures detaching from the frontal region. In the pycnocline, LCSs are able to detect filamentary structures that are not captured by the Eulerian fields.}

\begin{document}

\title{\textbf{\large{Three dimensional chaotic advection by mixed layer baroclinic instabilities}}}

\author{\textsc{Daniel Mukiibi,} \textsc{Gualtiero Badin,}
				\thanks{\textit{Corresponding author address:} 
Gualtiero Badin, Institute of Oceanography,
University of Hamburg,
Bundesstrasse 53,
D-20146 Hamburg,
Germany. 
				\newline{E-mail: gualtiero.badin@uni-hamburg.de}}\quad\textsc{Nuno Serra}\\
\textit{\footnotesize{Institute of Oceanography, University of Hamburg, Germany}}
}

%
\ifthenelse{\boolean{dc}}
{
\twocolumn[
\begin{@twocolumnfalse}
\amstitle

\begin{center}
\begin{minipage}{13.0cm}
\begin{abstract}
	\myabstract
	\newline
	\begin{center}
		\rule{38mm}{0.2mm}
	\end{center}
\end{abstract}
\end{minipage}
\end{center}
\end{@twocolumnfalse}
]
}
{
\amstitle
\begin{abstract}
\myabstract
\end{abstract}
\newpage
}

\section{Introduction}
\label{sec:submesoscales}
Observations \citep[e.g.][and references therein]{shcherbina_2015} and high resolution numerical modeling studies (e.g. \citet{leif_2008} and references therein) reveal the presence of a wide variety of ocean dynamical processes at scales smaller than the deformation radius, which have been referred to as submesoscale dynamics. 
Dynamics in this regime are characterized by Rossby ($\text{R}_\text{o}$) and bulk Richardson ($\text{R}_\text{i}$) numbers of $O$(1) \citep{leif_2008}, differing thus from dynamics at mesoscale and large scales, where $\text{R}_\text{o} << 1$ and $\text{R}_\text{i} >>1$.

One of the sources of submesoscale variability is given by mixed layer instabilities (MLIs) \citep{boccaletti_2007, fox-kemper_2008a}. Mixed layer (ML) fronts can be created, for example, by the passage of storms which leave areas of the ocean locally mixed \citep{price_1981, ferrari_2000}, by tidal mixing in the coastal regions \citep{badin_2009} and in upwelling regions where deeper, colder waters are brought to the surface \citep{calil_2010, bettencourt_2012}. ML fronts are dynamically unstable: after an initial geostrophic adjustment \citep{tandon_1994, tandon_1995,young_1994}, they undergo baroclinic instability, yielding ageostrophic MLIs with growth rates of the order of days \citep{haine_1998,molemaker_2005} and leading to ML restratification \citep{boccaletti_2007, fox-kemper_2008b}.  
The restratification of the surface ocean may  be further affected by other forms of instabilities such as symmetric instabilities \citep{haine_1998,taylor_2009}, while other dynamical factors like down-front wind stress have been found to slow down the restratification-mixing cycle of the upper ocean \citep{mahadevan_2010}. MLIs lead to the emergence of filamentary features. These filaments can create a form of nonlocal turbulence, in which the small scale motions are controlled by the large scale dynamics \citep[e.g.,][]{badin_2014, gula_2014}. Otherwise, the filaments can be formed by local frontogenesis, which takes the shape of elongated features \citep{ragone_2015}. The filaments are characterized by intensified relative vorticity, vertical velocity and strain rate \citep{mahadevan_2006b, leif_2008}. The filaments further undergo secondary instabilities (e.g. \citet{leif_2008, gula_2014}). The intensification of vertical velocities at submesocale has important effects on the budgets of buoyancy, mass and other tracers, for example facilitating the supply of nutrients and gases to the euphotic layers of the ocean thereby enhancing primary production in the ocean interior \citep{levy_2001}. Frontal dynamics can be important also for the transformation of water masses \citep{Thomas_Joyce_2010,badin_2010,badin_2013,Thomas_2013}. Further, MLIs might be able to penetrate in the underlying pycnocline where they might be important for the lateral mixing of tracers \citep{badin_2011}. 

The traditional techniques used in the definition and identification of coherent structures make use of Eulerian fields, defining them as localized, persisting regions with values of relative vorticity or strain rate larger than their surroundings \citep[e.g.,][]{calil_2010}. An alternative definition makes use of the Okubo-Weiss (OW) parameter, defined as the difference between the square of relative vorticity and horizontal strain \citep{okubo_1970, weiss_1991}. While the OW parameter sometimes correctly identifies coherent vortices \citep{boffetta_2001, harrison_2012}, and a strong correlation has been found to exist between zero level contours of the OW parameter and Lagrangian Coherent Structures (LCSs) \citep{dovidio_2009}, this technique is also observed to yield boundaries of vortices that are an underestimation of the actual sizes of the vortices \citep{haller_2000a, harrison_2012}. 
Further, and perhaps more seriously, the OW parameter is not an objective method to assess the flow coherence as it depends on the frame of reference in which the observations are made, and leads thus to an observer dependent assessment of flow coherency \citep{beron-vera_2013,haller_2015}.
In the current study, the OW parameter presents a further problem that is characteristic of ageostrophic instabilities: as stated previously, filamentary MLIs are characterized by intensified relative vorticity and strain rate in the same location, making the OW parameter an ill defined quantity.


Given these issues in studying chaotic stirring and in identifying the structures responsible for this stirring, in the current study, we concentrate on the Lagrangian approach to study the chaotic advection emerging from the MLIs using Finite Time Lyapunov exponents. 
%
 
Lyapunov exponents are defined in the asymptotic limit of infinite time intervals which renders them inapplicable to geophysical situations where velocity fields are only known for finite time intervals. As an alternative, Lyapunov exponents can be calculated for finite intervals of time, leading to the concept of Finite Time Lyapunov Exponents (FTLEs) \citep{haller_2000a, shadden_2005}. Differently from Lyapunov exponents defined on a strange attractor, FTLEs are not a global dynamical property of the flow and thus depend on the initial conditions of the calculated trajectories, i.e. on the initial position and on the initial time of release of the particles. This apparent limitation results however in the property of FTLEs being able to capture local features of the flow, such as hyperbolic regions and stirring/adiabatic mixing barriers \citep{lapeyre_2002}. Because the Lyapunov exponents define rates of exponential separation of particles (e.g. passive tracers), they become an important measure for the stirring and dispersive properties of the flow. The tendency of the flow to fill the chaotic region results in a nonlocal form of turbulence, suggesting that these features might provide the correct representation for submesoscale turbulence. The theory assumes that the velocity field prescribed by the flow is already known in form of analytic functions \citep[e.g.,][]{haller_2001, haller_2002, shadden_2005, lekien_2007, sulman_2013}, numerical simulations \citep[e.g.,][]{rypina_2007,rypina_2010,bettencourt_2012} or observation data taken by satellites \citep{beron_2008, waugh_2008, waugh_2012, harrison_2012}.  


Few studies have considered three dimensional FTLEs for geophysical flows due to the fact that such flows are predominantly two dimensional. Among the exceptions is the study by \citet{sulman_2013}, who considered the FTLEs and the resulting LCSs emerging from analytic 3D velocity fields. Their results show that appropriate approximations of 3D FTLEs should account for vertical shear of horizontal currents. In the present study, we consider a more geophysically relevant flow obtained from the instability of a ML front, in which the dynamics are dominated by the presence of stratification and rotation. The resulting instabilities are characterized by enhanced vertical velocities and vertical shear. We will thus  focus on the following questions: what is the chaotic stirring resulting from MLIs? What is the role of vertical velocities and vertical shear in determining the structure and magnitude of FTLEs? What are the differences between 3D and 2D FTLEs for MLIs? And, finally, how does the skeleton of MLIs turbulence, responsible for the chaotic stirring, look like? 

The manuscript will be arranged as follows: Section \ref{sec:theoretical_background} will provide the theoretical background on the computation of FTLEs.
Section \ref{sec:numerical_model} reports a brief description of the numerical model employed in the study. Section \ref{sec:methods} explains the methods used to obtain the particle trajectories from the velocity fields and the experiments performed in the study. Results obtained from the study and their discussion are given in Section \ref{sec:results}. In particular, in Section \ref{sec:results}a we show that, for ageostrophic MLIs, the OW parameter does not allow to identify the filamentary structures that are responsible for stirring: while the inability of the OW parameter to capture the LCSs and its lack of objectivity was already pointed by \citet{beron-vera_2013,haller_2015}, we show here an indeterminacy problem strictly linked to the ageostrophic character of MLIs; 
In Section \ref{sec:results}b we then characterize the MLIs using FTLEs. Different approximations are introduced in order to determine what controls the magnitude and spatial distribution of the FTLEs. The observation that 3D and 2D FTLEs differ for being controlled by vertical and horizontal shear of the currents, will be used in Section \ref{sec:results}c to derive a scaling for these quantities; Section \ref{sec:results}d analyses the characteristics of the resulting FTLE fields, such as their probability distribution functions (PDFs) and the power spectra in the different approximations, in particular discussing the local and nonlocal nature of the emerging turbulence at different scales; The analysis is concluded in Section \ref{sec:results}e determining the skeleton of MLIs turbulence, responsible for the chaotic stirring. 
Finally, Section \ref{sec:conclusions} reports the conclusions and gives final remarks.

\section{Theoretical background of FTLEs}
\label{sec:theoretical_background}
\subsection{Calculation of FTLEs}
Consider the velocity field of a flow described by the first order system of ordinary differential equations
\begin{equation}
\frac{\text{d}}{\text{dt}}{\bf x} = \text{\bf u}({\bf x},t)~,
\label{eq:dynamical_system}
\end{equation}
where ${\bf x} = (x,y,z)$ are the three dimensional particle trajectories. The perturbation to a particle trajectory ${\bf x}(t)$ in the time interval $[t_1,t_2]$ is computed as  $\boldsymbol \delta (t_2) = {\bf x}(t_2) - {\bf x}(t_1)$. The velocity field $\text{\bf u}$ can thus be considered as a map of the flow $\phi$ which takes the initial position of the particle ${\bf x}(t_1)$ and returns the final position ${\bf x}(t_1+t_2)$ of the particle at a later time $t_1+t_2$, 
\begin{equation}
\phi_{t_1}^{t_2}|{\bf x}(t_1)\rangle =\left|{\bf x}(t_1+t_2) \right\rangle ~.
\label{eq:bra_ket}
\end{equation}
where a bra-ket notation has been adopted. 
Using a Taylor expansion about $|{\bf x}(t_1)\rangle$, 
a perturbation $\boldsymbol \delta(t_1)$ to a particle trajectory ${\bf x}(t_1)$ is evolved linearly by the flow map as 
 \begin{equation}
 \phi_{t_1}^{t_2}|{\bf x}(t_1)+{ \boldsymbol \delta}(t_1)\rangle= \phi_{t_1}^{t_2}|{\bf x}(t_1)\rangle + 
{ \boldsymbol \delta}(t_1) \frac{{\text d}}{{\text d\bf{x}}} \phi_{t_1}^{t_2}|{\bf x}(t_1)\rangle + O(\delta^2)~.
 \label{eq:expansion}
 \end{equation}
 Assuming that the flow map defined by (\ref{eq:expansion}) is at leading order linear,  the equation for the evolution of the perturbation of a particle trajectory is 
 \begin{equation}
 \phi_{t_1}^{t_2}|{ \boldsymbol \delta}(t_1)\rangle= |{ \boldsymbol \delta}(t_1+t_2)\rangle={ \boldsymbol \delta}(t_1) \dfrac{{\text d}}{{\text d{\bf x}}} \phi_{t_1}^{t_2}\left|{\bf x}(t_1)\right\rangle ~,
 \label{eq:perturbation}
 \end{equation}
and its square norm is
 \begin{eqnarray}
  \left\|{ \boldsymbol \delta}(t_1+t_2)\right\|^2 = \left\langle{ \boldsymbol \delta}(t_1+t_2) |{ \boldsymbol \delta}(t_1+t_2)\right\rangle  \\ \nonumber
  =  \left\langle{ \boldsymbol \delta}(t_1)\frac{\text d}{\text {d{\bf x}}}\phi_{t_1}^{t_2}{\bf x}(t_1)|{ \boldsymbol \delta}(t_1)\frac{\text d}{\text {d{\bf x}}}\phi_{t_1}^{t_2}{\bf x}(t_1)\right\rangle ~,
 \label{eq:perturbation_norm1}
 \end{eqnarray}
where $ || \cdot || $ is the three dimensional Euclidean norm.
Thus, the square of the norm of the resulting perturbation in a particle trajectory after a time $(t_1+t_2)$ is given by the expression 
\begin{equation}
\|{ \boldsymbol \delta}(t_1+t_2)\|^2 = \langle { \boldsymbol \delta}(t_1)|\left(\frac{\text d}{\text {d{\bf x}}} 
\phi_{t_1}^{t_2}{\bf x}(t_1)\right)^\dagger\left(\frac{\text d}{\text {d{\bf x}}} 
\phi_{t_1}^{t_2}{\bf x}(t_1)\right)|{ \boldsymbol \delta}(t_1)\rangle ~,
\label{eq:perturbation_norm2}
\end{equation}
where $\left(\cdot \right)^\dagger$ is obtained by taking the complex conjugates of the entries of the 
matrix $\left( \cdot \right)$ and then taking its transpose.  The matrix given by 
\begin{equation}
{\boldsymbol \Delta}\left({\bf x}(t_1), t_1,t_2 \right) =\left(\frac{\text d}{\text {d{\bf x}}} 
\phi_{t_1}^{t_2}{\bf x}(t_1)\right)^\dagger\left(\frac{\text d}{\text {d{\bf x}}} \phi_{t_1}^{t_2}{\bf x}(t_1)\right)~, 
\label{eq:deformation_tensor}
\end{equation} 
is known as the finite time Cauchy-Green deformation tensor. From its construction, $\boldsymbol \Delta$ is a real positive definite tensor, with real eigenvalues.
Equation (\ref{eq:perturbation_norm2}) can therefore be written as 
 \begin{eqnarray}
\|{ \boldsymbol \delta}(t_1+t_2)\|^2 &=& \left\langle { \boldsymbol \delta}(t_1)|{ \boldsymbol \Delta}\left({\bf x}(t_1),t_1,t_2 \right) |{ \boldsymbol \delta}(t_1)\right\rangle \\ \nonumber
&=& C\langle { \boldsymbol \delta}(t_1)|{ \boldsymbol \delta}(t_1)\rangle= C\|{ \boldsymbol \delta}(t_1)\|^2 ~,
\label{eq:norms}
\end{eqnarray}
where $C$ is the eigenvalue of the operator ${ \boldsymbol \Delta}\left({\bf x}(t_1), t_1,t_2 \right)$ and is defined such that it satisfies the relation  ${\boldsymbol \Delta}\left({\bf x}(t_1), t_1,t_2 \right) |{ \boldsymbol \delta}(t_1)\rangle= C |{ \boldsymbol \delta}(t_1)\rangle$. 

In a chaotic advection flow regime, initially infinitesimal perturbations in particle paths grow exponentially
 i.e $|| \boldsymbol\delta(t_1+t_2)||=|| \boldsymbol\delta(t_1)|| \exp \left[ \lambda (t_2-t_1) \right]$, where the scalar $\lambda$ is the Finite Time Lyapunov Exponent (FTLE) \citep{haller_2000a,haller_2000b, haller_2001,shadden_2005, lekien_2007}. The FTLEs can thus be calculated from the expression
\begin{equation}
\lambda = \frac{1}{(t_2-t_1)} \text{log}\left( \frac{|| \boldsymbol\delta(t_1+t_2)||}{|| \boldsymbol\delta(t_1)|| }\right) = \frac{1}{(t_2-t_1)}\text{log}\left(C_{max} \right)^{1/2}~,
\label{eq:lambda1}
\end{equation} 
where $C_{max}$ is the largest of the eigenvalues of the operator $\boldsymbol \Delta$ defined in (\ref{eq:deformation_tensor}). The eigenvector associated to $C_{max}$ corresponds to the direction along which maximum separation of initially infinitesimally close particles occurs. Equation (\ref{eq:lambda1}) shows that the scalar field $\lambda$ is a measure of the rate of particle separation in the time interval $[t_1,t_2]$. 
Equation (\ref{eq:lambda1}) also shows that the length of the time interval of integration $[t_1,t_2]$ determines the magnitude of the FTLEs following an inverse relation. Longer integration times yield finer and more detailed FTLE fields \citep[e.g.,][]{lapeyre_2002, shadden_2005, harrison_2012}. However, from a geophysical point of view, it is also important to select the length of the time interval
of integration based on the flow dynamics. A meaningful time interval should be long enough to cover the life span of the  longest dynamics in the flow domain, ensuring that all the stirring influences of vortices and filaments are fully captured in the calculation of the FTLEs.

\section{Numerical model}
\label{sec:numerical_model}
A ML front in a channel configuration is here considered, using a numerical primitive equation model, the Massachusetts Institute of Technology general circulation model (MITgcm), in hydrostatic mode \citep{mitgcm_1997b, mitgcm_1997a}. A similar model configuration as in \citet{boccaletti_2007} is adopted. The domain spans 192 km both in the zonal and meridional directions and is 300 m deep. The zonal and meridional resolutions are both set at 500 m. The vertical resolution is uniformly set as 5 m. The channel is  re-entrant with periodic boundary conditions along the zonal direction. The meridional walls of the channel  are rigid and impermeable, with free slip boundary conditions. The bottom of the channel is set with no topography and with free slip boundary conditions. The top of the channel satisfies free surface boundary conditions. Model parameters used in the numerical simulations are presented in Table \ref{tab:model_parameters}. The channel is initialized with a ML front with a density contrast aligned in the zonal direction and 100 m deep. The ML front is positioned 96 km north of the southern boundary of the channel. The southern part of the channel contains lighter warm and more saline waters at the surface, while  the northern part is initialized with heavier, cold waters at the surface. The ML lies upon an initially quiescent pycnocline with flat isopycnals. The temperature and salinity profiles used in the reference simulation set an initial uniform buoyancy frequency in the ML which, following an hyperbolic tangent function, decreases with depth in the pycnocline. The dynamically unstable ML front is then allowed to adjust without any restoration. 

It should be noted that, as the model is based on primitive equations, the vertical velocity is only diagnosed from the divergence of the horizontal velocities. However, for the set-up and scales analyzed in this study, the most important part of the vertical velocity is captured by the divergence of the horizontal flow. See, e.g. \citet{mahadevan_2006b} and \citet{mahadevan_2006a}.

\section{Methods}
\label{sec:methods}
\subsection{Computation of particle trajectories}
A time interval $\tau=t_2 - t_1$ during which the particle motion is investigated is selected. The lower limit $t_1$ is selected at an instant after the initial spin-up of the model, when the flow is well developed to reveal the stirring influence of the instability. The value of $t_2$ is made as large as possible depending on the computational resources available, but less than the time at which the instabilities reach the meridional boundaries of the channel. The velocity field in the time window $\tau$ is then written out every 15 minutes. A regular grid of particles is set at each grid point in the domain, for a total of 8,609,516 particles. 
The particle trajectories are integrated using a Runge-Kutta fourth order scheme. For spatial interpolations of the velocity field, a tricubic scheme is adopted while a linear scheme is used for temporal interpolations. Computation of trajectories is not considered for particles on the boundaries of the channel. FTLEs are calculated using both forward and backward integration in time, where the backward integration is performed in the interval $\left[ t_2,t_1 \right]$. A note of caution is here obligatory: the forward and backward integration allows to use the same flow, however it relies on different initial conditions. This choice has been made in order to compare the statistics of the FTLEs, however no comparison of snapshots of the field should be attempted. 

\subsection{Computation of FTLEs}
In the current study, we consider the operator, $\frac{\text d}{\text {d{\bf x}}} \phi_{t_1}^{t_2}{\bf x}(t_1)$ 
to be the $3 \times 3$ matrix {\bf D} whose entries are numerically obtained as finite differences. For a particle located away from the channel boundaries, there are six nearest neighbors, laying along the three cardinal directions i.e North (N) - South (S), East (E) - West (W) and Top (T) - Bottom (B) (Fig. \ref{fig:convergence_of_ftles}a). Components of the deformation tensor are computed as  

\begin{eqnarray}
{\bf D}= 
\begin{pmatrix}
\left(\dfrac{x_2^E-x_2^W}{x_1^E - x_1^W}\right)& \left(\dfrac{x_2^N-x_2^S}{y_1^N-y_1^S}\right)&\left(\dfrac{x_2^T-x_2^B}{z_1^T-z_1^B}\right) \\
\left(\dfrac{y_2^E-y_2^W}{x_1^E - x_1^W}\right)  & \left(\dfrac{y_2^N-y_2^S}{y_1^N-y_1^S}\right)& \left(\dfrac{y_2^T-y_2^B}{z_1^T-z_1^B}\right) \\
\left(\dfrac{z_2^E-z_2^W}{x_1^E - x_1^W}\right)  & \left(\dfrac{z_2^N-z_2^S}{y_1^N-y_1^S}\right)& \left(\dfrac{z_2^T-z_2^B}{z_1^T-z_1^B}\right)
\end{pmatrix}~,
\label{eq:matrix_approx}
\end{eqnarray}
where ${\bf x}_1={\bf x}(t_1)$ and ${\bf x}_2={\bf x}(t_2)$ are the particle positions. The FTLEs $\lambda$ are then obtained from (\ref{eq:lambda1}), where
$C_{max}$ is the maximum of the eigenvalues of $\left(\text{\bf D}^\text{T}\text{\bf D}\right)$.

\subsection{Numerical experiments}
A set of five numerical experiments have been conducted with different values of the initial surface density contrast $\Delta\rho$ (Table \ref{tab:experiments}). For a ML of depth $H_{ML}$, the deformation radius can be estimated from the relation $R_{d}=M^2 H_{ML} / f^2$, where, for a ML front aligned along the zonal direction, $M^2=b_y$ is the buoyancy gradient across the front, with the buoyancy $b=-g\Delta\rho/\rho_s$, where $g$ is the gravitational acceleration and $\rho_s$ is the reference density. In the pycnocline, the deformation radius is calculated as $R_d=N_{max} H_{tot} /f$, where $N_{max}$ is the maximum value of the buoyancy frequency and $H_{tot}$ is the channel depth. Since the resulting instabilities in each of the experiments have different growth rates and deformation radii, the time window used to calculate the FTLEs (Table \ref{tab:experiments}) differs accordingly to the time required for the instabilities to reach the meridional boundaries of the channel. The experiment with $\Delta\rho = 0.4~ \text{kg m}^{-3}$ is taken as the reference experiment. 

To investigate the contribution of the various components of the deformation tensor $\text{\bf D}$ to $\lambda$, four realizations of $\text{\bf D}$ are considered. To investigate the role of vertical velocities, the vertical displacement terms $\partial z_2/{\partial x_1}$ and ${\partial z_2}/{\partial y_1}$ are set to zero, leading to 
\begin{eqnarray}
{\bf D_1}({\bf x},t_1,t_2) = 
 \begin{pmatrix}
    \dfrac{\partial x_2}{\partial x_1}& \dfrac{\partial x_2}{\partial y_1} & \dfrac{\partial x_2}{\partial z_1}\\
    \dfrac{\partial y_2}{\partial x_1}& \dfrac{\partial y_2}{\partial y_1} & \dfrac{\partial y_2}{\partial z_1}\\
    0& 0 & 1
  \end{pmatrix}~.
   \label{eqn:cauchy_2}
  \end{eqnarray}
To deduce the contribution of vertical shear to the overall rate of particle separation, the terms ${\partial x_2}/{\partial z_1}$ and ${\partial y_2}/{\partial z_1}$ are set to zero yielding 
 \begin{eqnarray}
{\bf D_2}({\bf x},t_1,t_2) = 
 \begin{pmatrix}
    \dfrac{\partial x_2}{\partial x_1}& \dfrac{\partial x_2}{\partial y_1} & 0\\
    \dfrac{\partial y_2}{\partial x_1}& \dfrac{\partial y_2}{\partial y_1} & 0\\
    \dfrac{\partial z_2}{\partial x_1}& \dfrac{\partial z_2}{\partial y_1} & 1
  \end{pmatrix}~.
  \label{eqn:cauchy_2}
  \end{eqnarray}
Setting the joint contribution of vertical displacements and vertical shear to zero, yields a reduction to a two dimensional system in which particle separation is effected only by the horizontal strain
  \begin{eqnarray}
{\bf D_3}({\bf x},t_1,t_2) = 
 \begin{pmatrix}
    \dfrac{\partial x_2}{\partial x_1}& \dfrac{\partial x_2}{\partial y_1} & 0\\
    \dfrac{\partial y_2}{\partial x_1}& \dfrac{\partial y_2}{\partial y_1} & 0\\
    0& 0 & 1
  \end{pmatrix}~.
   \label{eqn:cauchy_3}
  \end{eqnarray}
Finally, setting the horizontal strain and vertical displacement terms to zero, yields 
\begin{eqnarray}
{\bf D_4}({\bf x},t_1,t_2) = 
 \begin{pmatrix}
    1&0 & \dfrac{\partial x_2}{\partial z_1}\\
    0& 1 & \dfrac{\partial y_2}{\partial z_1}\\
    0& 0 & 1
  \end{pmatrix}~,
   \label{eqn:cauchy_4}
  \end{eqnarray}
from which the contribution of vertical shear to particle separation is determined.
The resulting FTLE approximations from the above approximations of the Cauchy-Green deformation tensor will be denoted as follows
 \begin{eqnarray}
 \text{3D} &=& \frac{1}{2|\tau|}\text{log}C~,
\label{eq:full} \\
 \text{approx1} &=& \frac{1}{2|\tau|}\text{log}C_1~,
 \label{eq:approx1} \\
 \text{approx2} &=& \frac{1}{2|\tau|}\text{log}C_2~,
  \label{eq:approx2} \\
 \text{approx3} &=& \frac{1}{2|\tau|}\text{log}C_3~,
 \label{eq:approx3}\\
 \text{approx4} &=& \frac{1}{2|\tau|}\text{log}C_4~,
 \label{eq:approx4}
 \end{eqnarray}
 where $C$, $C_1$, $C_2$, $C_3$ and $C_4$ are respectively the maximum of the eigenvalues of the operators $\text{\bf D}^\text{T}\text{\bf D}$, $\text{\bf D}_1^\text{T}\text{\bf D}_1$, $\text{\bf D}_2^\text{T}\text{\bf D}_2$, $\text{\bf D}_3^\text{T}\text{\bf D}_3$ and $\text{\bf D}_4^\text{T}\text{\bf D}_4$. The absolute value $\left(| \cdot |\right)$ of $\tau$ in (\ref{eq:full}) - (\ref{eq:approx3}) is emphasized since the sign of $\tau$ changes from being positive for forward FTLEs to negative for backward FTLEs. Fig. \ref{fig:convergence_of_ftles}b,c show the variation of area averages of FTLEs with the integration time $\tau$ for the 3D and approx2 respectively. The integrated values of the FTLEs are observed to converge at all depths in about 470 hours, corresponding to $\sim 19.6$ days. \citet{badin_2011} reported that in this time, the separation of passive tracer was still exponential and thus in a chaotic advection regime. As we are interested in the statistical properties of stirring, using a shorter interval would yield a large change in the shape of the PDFs and the spectra for small changes in the interval length, while with this choice, the statistics appear to be quasi-stationary, in the limits of the time evolving flow associated with the freely decaying front.
 
 \section{Results}
\label{sec:results}
\subsection{Eulerian fields}
At the surface (Fig. \ref{fig:eulerian_fields}a,b), the MLIs are visible in the form of filaments along which the relative vorticity and strain rate are intensified. Isolated vortices which break away from the main frontal regions are observed as regions with large vorticity cores, surrounded by high values of strain rate. For example, a dipolar structure is observed in the lower left corner of the domain. While the structure appears to be an isolated vortex, closer inspection, changing for example the range of the color bar, allows to recognize its dipolar nature. In the channel interior, the filamentary structures disappear leaving larger regions with enhanced values of vorticity and strain rate (Fig. \ref{fig:eulerian_fields}d,e). The existence of regions of enhanced vorticity and strain rates in the interior, confirms previous observations that MLIs can penetrate into the ocean interior \citep{badin_2011}. The OW parameter identifies isolated vortices as vorticity dominated cores surrounded by regions of high strain rate. Filamentary structures are however difficult to characterize from the OW parameter field, since both their vorticity and strain rate are intensified, yielding regions with alternating positive and negative values of the OW parameter (Fig. \ref{fig:eulerian_fields}c). One example is given by the surface ageostrophic filament extending at $x \sim 40~km$ and $y\sim 40-90~km$, which has a strong signature in both the vorticity and strain rate fields, but that disappears in the OW field (Fig. \ref{fig:eulerian_fields}a,b,c). The failure to detect filaments by the Eulerian fields is a motivation for the choice to adopt a Lagrangian approach in studying the stirring properties of MLIs.

\subsection{FTLE fields}
The forward 3D FTLE fields (Fig. \ref{fig:FTLEs}) show a much more complex structure than the Eulerian fields (Fig. \ref{fig:eulerian_fields}) at all depths. Isolated vortices are characterized by high values of FTLEs on both their interior and  boundaries. The reason for the presence of regions with high values of FTLEs within the vortices is due to the unbalanced nature of the vortices, which have a spiraling structure associated to the divergence of the flow, resulting in a fine FTLEs structure also in their interior. Filaments in the main frontal region are instead characterized by regions with high values of FTLEs  alternating with regions of low values of FTLEs in a very fine structure. This shows that in the frontal region, characterized by an interplay of MLIs and their filamentary structures, secondary instabilities act to fold, stretch and entangle the Lagrangian structure of turbulence. Eventually, for times longer than the integration time used, the FTLEs would merge to create a chaotic region.  
Noticeable, stirring is much more complex than revealed by Eulerian measures.  3D FTLEs are finer at the ML base than at the surface (Fig. \ref{fig:FTLEs}b), with filaments and vortex boundaries with a more distinct appearance. In the channel interior, filamentary structures are detected by the FTLE field in locations where the Eulerian fields are rather featureless (Fig. \ref{fig:FTLEs}c). The different appearance of the FTLEs at the sea surface from the FTLEs at base of the ML and in the interior is related to the fact that at depth the flow is weaker and thus acts to tangle less the FTLEs, with the entanglement decreasing at depth with the strength of the flow.

\begin{figure*}[htb]
\centering
\begin{tabular}{@{}c@{}}   
\noindent\includegraphics[height=0.51\textheight,width=0.8\textwidth]{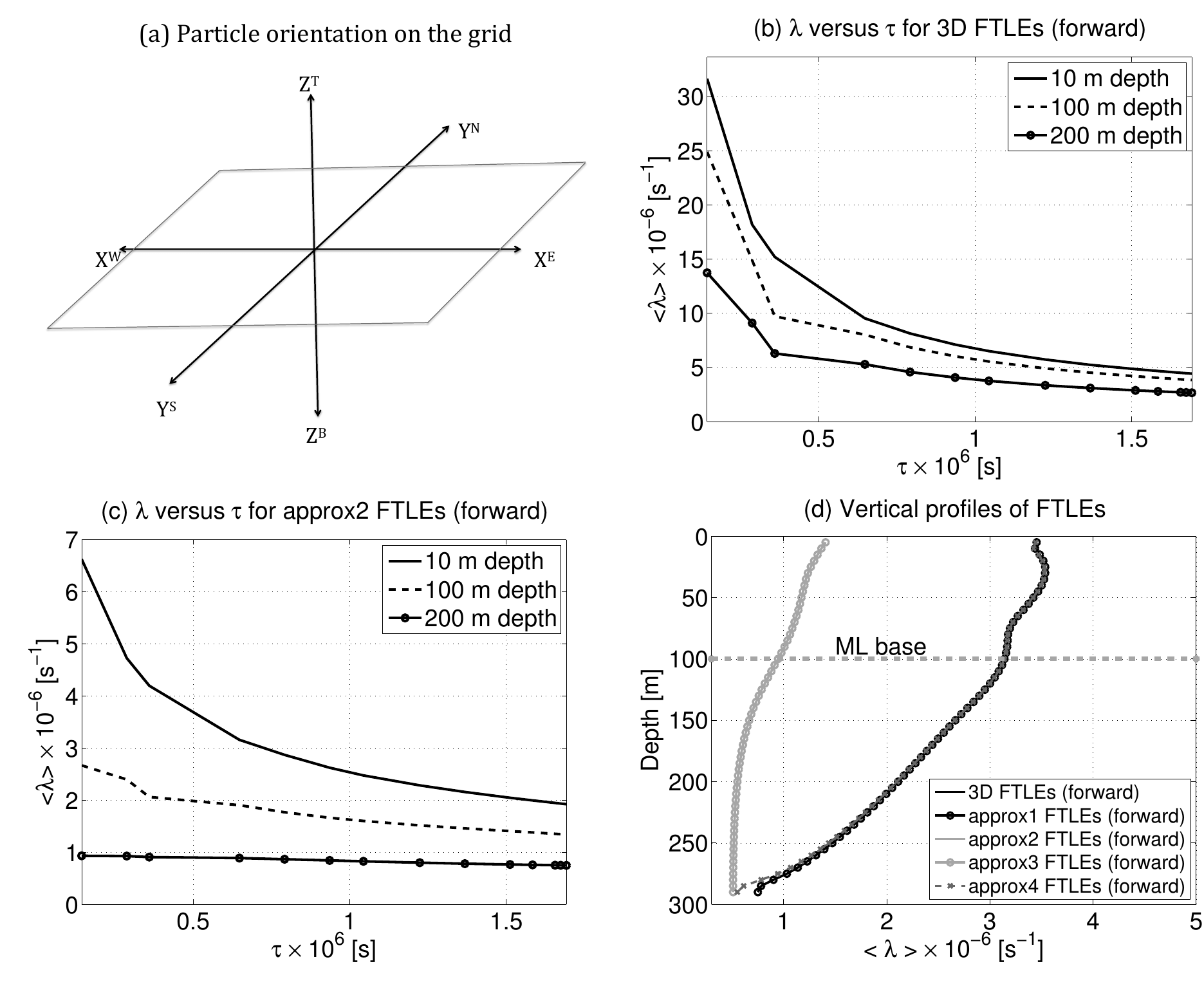}
 \end{tabular}
 \caption{(a) Particle positions on the model grid. Each particle has six nearest neighbors aligned along each of the cardinal directions. (b{)} Time evolution of the area averages of 3D FTLEs at 10 m (continuous line), 100 m (dashed line) and 200 m (dot dashed line) in the reference simulation. (c{)} Time evolution of approx2 FTLEs at 10 m (continuous line), 100 m (dashed line) and 200 m (dot dashed line) in the reference simulation. (d) Vertical profiles of the averaged FTLEs for the different approximations of FTLEs in the reference simulation at day 60. }
  \label{fig:convergence_of_ftles}
\end{figure*}

\begin{figure*}[htb]
\centering
\begin{tabular}{@{}c@{}}
\noindent\includegraphics[height=0.75\textheight,width=0.9\textwidth]{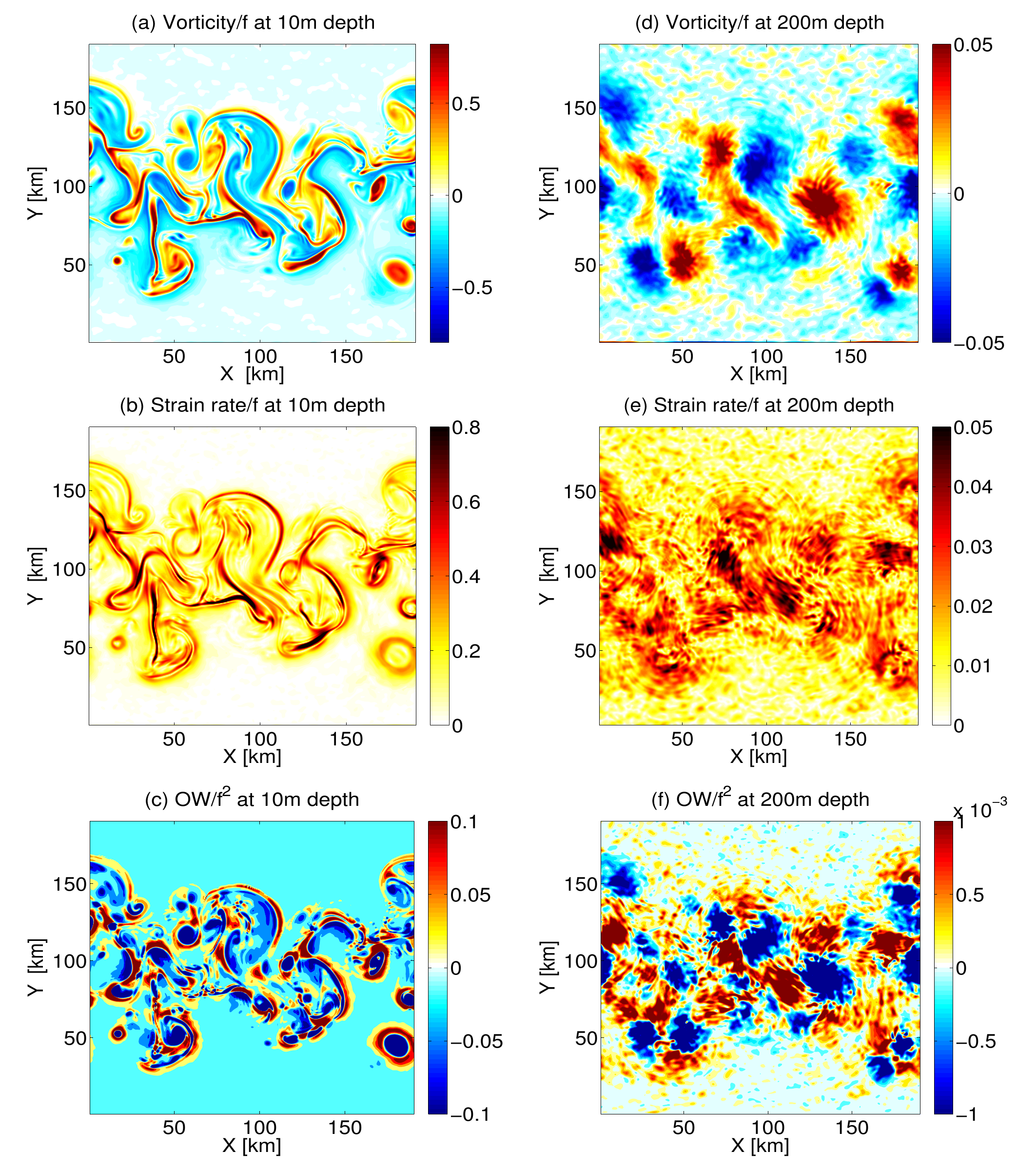}                                              
\end{tabular}
\caption{Eulerian fields evaluated at day 60 of the reference simulation. {Left column}: normalized (a) relative vorticity, (b) strain rate and {(}c) OW parameter at 10 m depth. {Right column}: normalized (d) relative vorticity,  (e) strain rate and (f) OW parameter  at 200 m depth.} 
  \label{fig:eulerian_fields}
\end{figure*}

\begin{figure*}[htb]
\centering
\begin{tabular}{@{}c@{}}                                                   
\noindent\includegraphics[height=0.75\textheight,width=0.9\textwidth]{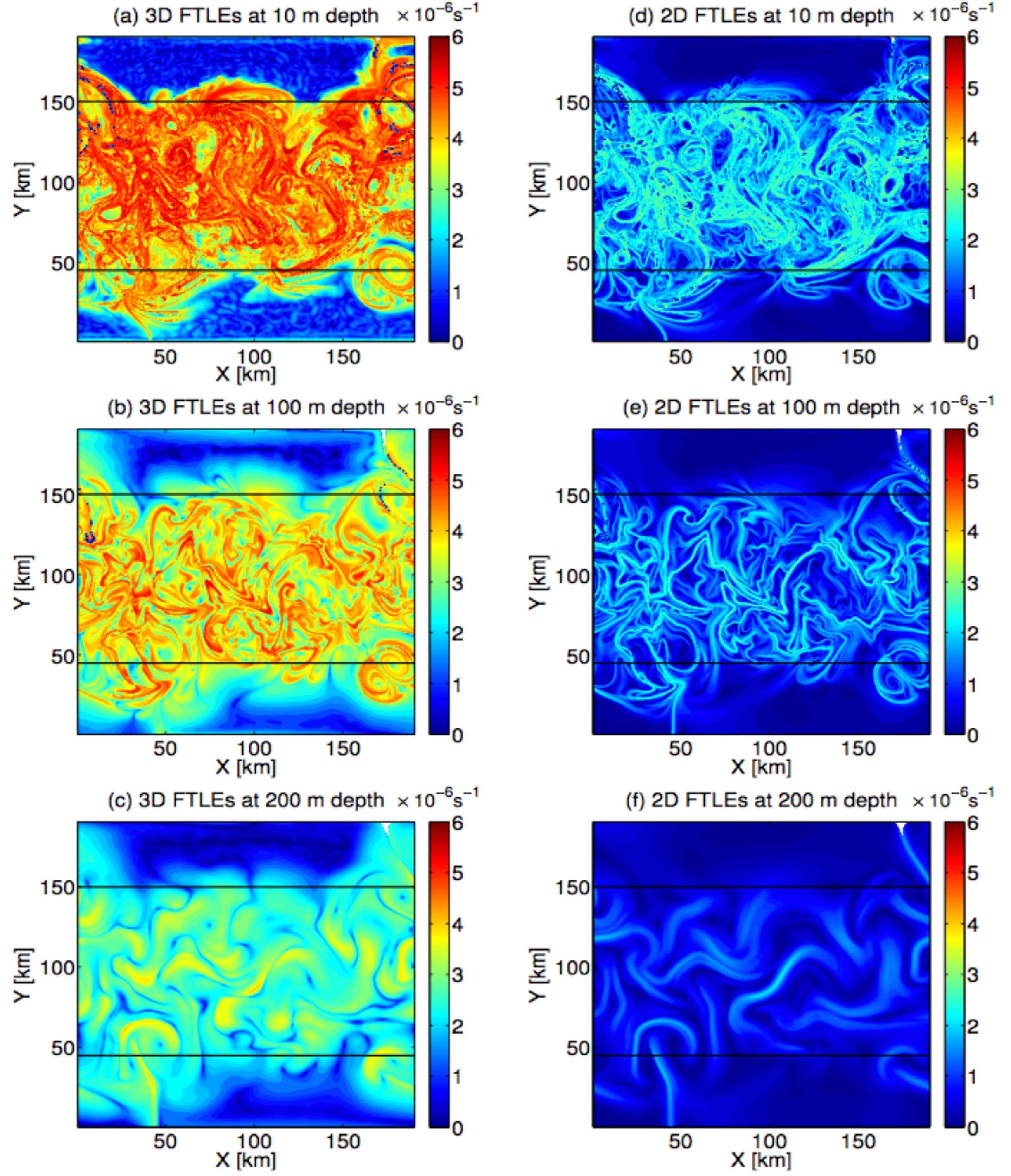}
\end{tabular}
 \caption{ Forward 3D FTLEs (left column) and forward 2D FTLEs (right column) at depths of 10 m (a,d), 100 m (b,e) and 200 m (c,f) in the reference simulation at day 60. The black horizontal lines demarcate the region for which further analysis of FTLEs is considered.}
  \label{fig:FTLEs}
\end{figure*}

The horizontally averaged 3D FTLEs (Fig. \ref{fig:convergence_of_ftles}d, black line) show that the 3D FTLEs have larger values in the ML, with a local maximum in the middle of the ML, in agreement with the observation from numerical simulations that MLIs produce stronger fluxes in the middle of the ML \citep{fox-kemper_2008a}, and have a fast decrease below the ML base, showing however non zero values at all depths.
Analysis of the vertical structure of horizontally averaged forward FTLEs from the different approximations (Fig. \ref{fig:convergence_of_ftles}d) shows that 3D (black line), approx1 (black dotted line) and approx4 (gray dot dashed line) FTLEs are indistinguishable at all depths. The same result holds for approx2 (gray line) and approx3 (gray dotted line) FTLEs, which are coincident at all depths, indicating that the vertical velocity does not play a significant role in determining the size of FTLEs, but that the magnitude of the FTLEs is dominated by the vertical shear. The analysis of the vertical structure of horizontally averaged FTLEs from the different approximations for the backward integration (not shown) yields the same results as the forward integration.

Due to the coincidence of the 3D, approx1 and approx4, as well as of approx2 and approx3 FTLEs, in the remaining only the results from 3D and approx2 FTLEs will be presented, with the approx2 FTLEs henceforth referred to as 2D FTLEs.  

2D FTLEs show ridges, which in first approximation are defined as local maxima (and minima of the negative) of the FTLE field, in the same location of the ridges of the 3D FTLEs field (Fig. \ref{fig:FTLEs}d,e,f). The ridges found for the 3D and 2D cases are in the same location as they are associated to the local intensification of vertical shear and horizontal strain, which are in turn associated to the localized ageostrophic instabilities.
Note that the ridges of the FTLEs do not denote LCSs, as it is now recognised that ridges have non zero flux across them \citep{haller_2015}. The values of the 2D FTLEs are however about half of the values of the 3D FTLEs. Further, the 2D FTLEs seem to show a smaller degree of entanglement of the FTLE field in the frontal region.
The large difference in the magnitude of FTLEs along locations of maximal and weak stretching of fluid patches yields well defined FTLE fields at all depths. Vortex boundaries, narrow regions separating dipoles of vortices and frontal structures are characterized by large values of FTLEs (Fig. \ref{fig:FTLEs}f). The vertical profiles of 2D FTLEs reveal that in addition to only being approximately half the values of 3D FTLEs,  2D FTLEs are surface intensified while their values quickly decrease below the ML (Fig. \ref{fig:convergence_of_ftles}d). This surface intensification of 2D FTLEs is also revealed by the observation that, for all $\tau$, the difference between the area averaged 2D FTLEs at different depths are larger than the difference between the area averaged 3D FTLEs at different depths (Fig. \ref{fig:convergence_of_ftles}b,c). At $200$ m depth, the values of 2D and 3D FTLEs have reduced by $\sim 80 \%$ and $\sim 40\%$ of their respective values at the ML base (Fig. \ref{fig:convergence_of_ftles}d). The slow decrease of 3D FTLE values from the base of the ML to the channel interior, shows that the vertical shear is able to sustain high rates of particle separation at depth. 

\begin{figure*}[htb]
\centering
\begin{tabular}{@{}c@{}}                                                   
\noindent\includegraphics[height=0.55\textheight,width=0.8\textwidth]{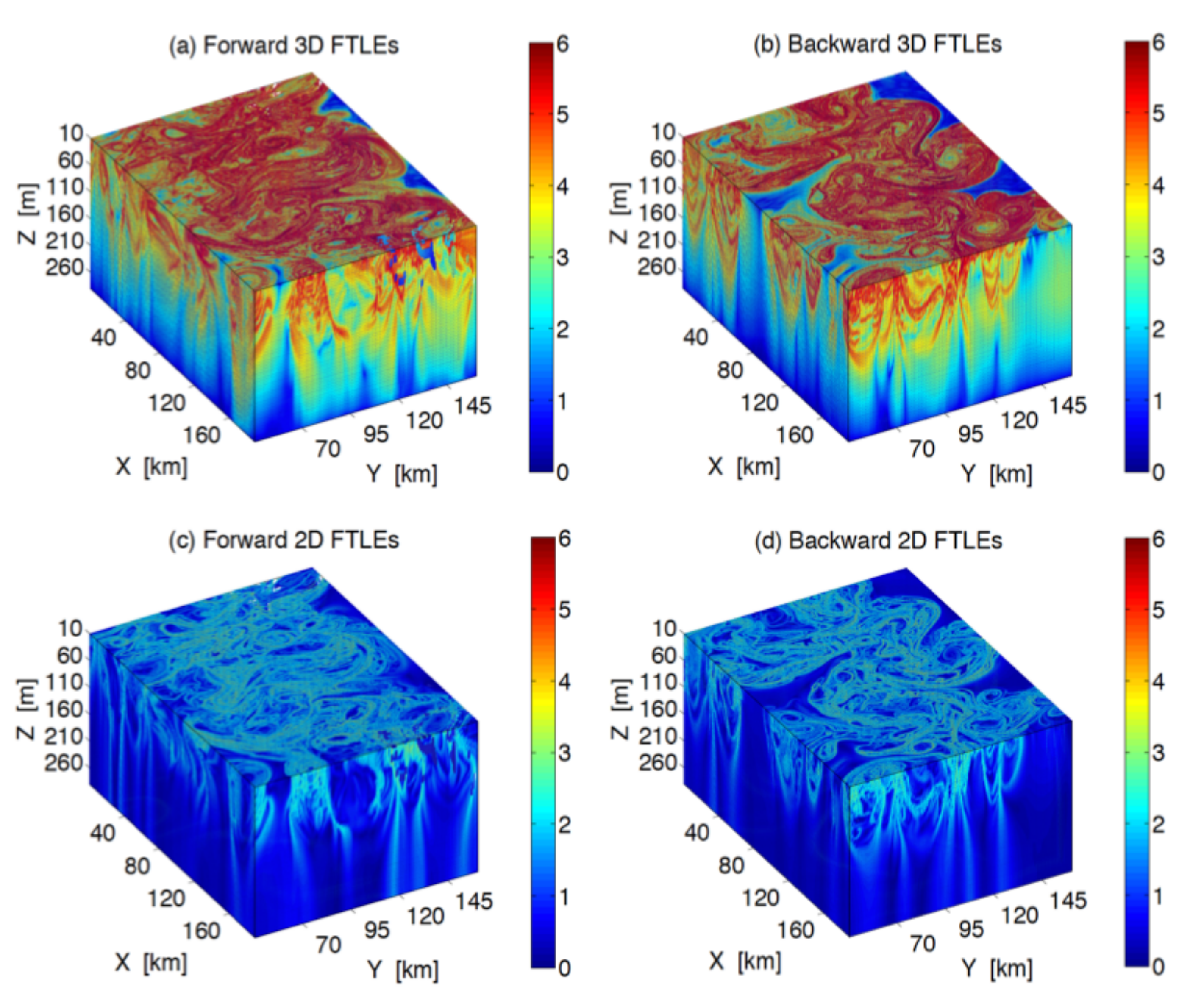}
\end{tabular}
 \caption{Left column: Forward in time (a) 3D and (c{)} 2D FTLEs. Right column: Backward in time  (b) 3D and (d) 2D FTLEs. All quantities have units of $10^{-6}\text{s}^{-1}$. Only the region shown between black lines in Fig.  \ref{fig:FTLEs} is presented.}
  \label{fig:3d_FTLEs}
\end{figure*}

3D FTLEs are thus able to "penetrate" in the channel interior, filling the volume of the channel (Fig. \ref{fig:3d_FTLEs})
where they show curtain-like structures that form the template for stirring in the channel. These curtain-like structures have also been found in previous studies that have considered 3D \citep{lekien_2007} and quasi 3D velocity fields \citep{bettencourt_2012}. 
Further, area averages of forward in time FTLEs are found to exhibit values comparable to their corresponding backward in time FTLE approximations at all depths (not shown).
It should be noted however that the forward and backward FTLEs have been calculated using different initial conditions, so no comparison between the backward FTLEs, which are calculated in the time interval $\left[t_2,t_1\right]$, and the Eulerian fields, which are defined at time $t_1$, should be attempted.
    
The relationship between the local value of the FTLEs and the vertical shear suggests the existence of a scaling relationship between the two quantities, which will be studied next. 

\subsection{Scaling relationship between the FTLEs and the vertical shear}
\label{subsec:lambda_relation}
\begin{figure*}[htb]
\centering
\begin{tabular}{@{}c@{}}      
\noindent\includegraphics[width=0.5\textwidth]{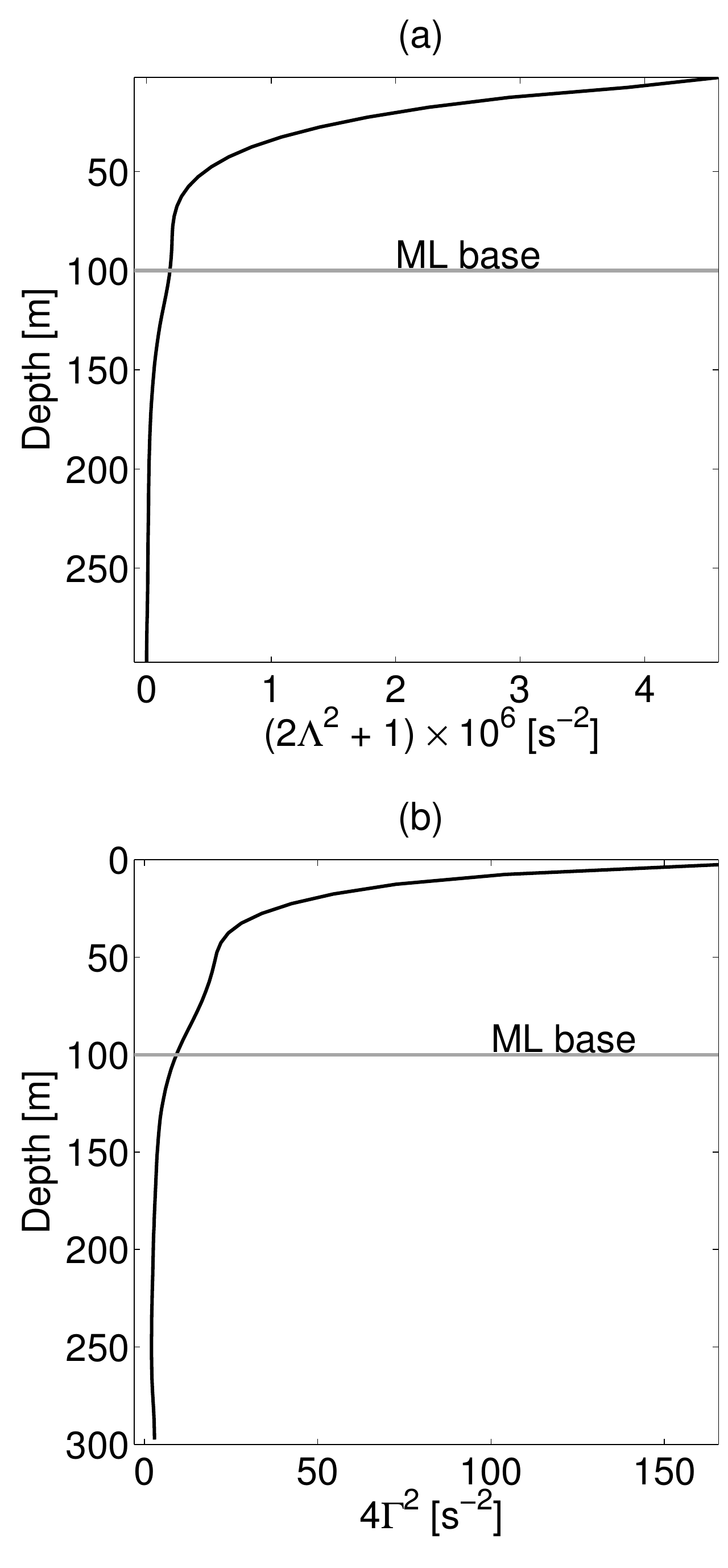}
\end{tabular}
 \caption{Vertical profiles of the area averaged (a) $2 \Lambda^2 +1$ and (b) $4 \Gamma^2$ in the reference simulation at day 60. As $2 \Lambda^2 +1 > 4 \Gamma^2$ at all depth, $\log_{4 \Gamma^2} \left( 2 \Lambda^2 +1 \right) > 1$ and $\lambda_\text{3d} > \lambda_\text{2d}$. }
 \label{fig:figure4b}
\end{figure*}

\noindent Consider a system in geostrophic and hydrostatic balance, so that the thermal wind relation 
\begin{equation}
 ~ \frac{\partial \boldsymbol U_g}{\partial z} =  \frac{g}{f \rho_{s}} \hat{k} \times \nabla \rho~,
\label{eq:thermal_wind}
\end{equation} 
holds, where $\boldsymbol U_g$ is the geostrophic current.
Approximating the derivatives using finite differences, (\ref{eq:thermal_wind}) yields 
\begin{equation}
\Lambda_i = \frac{\Delta {{\boldsymbol U}_g}}{\Delta z} \Delta t = \frac{g \Delta t}{f \rho_{s} \Delta x_i}\left(\Delta\rho\right) ~,
\label{eq:shear}
\end{equation} 
where $\Delta t$ is the time step of integration for the particle trajectories. 
Further, considering the flow gradient tensor in general terms as 
\begin{eqnarray}
   \begin{pmatrix}
    \Gamma_1& \Gamma_2& \Lambda_1\\
    \Gamma_3& \Gamma_4& \Lambda_2\\
    \omega_1& \omega_2 & 1
  \end{pmatrix}~,
  \label{eq:flow_gradient_tensor}
  \end{eqnarray}
  where $\Gamma_i$ and $\omega_i$ are the components of the horizontal shear of the horizontal and vertical currents multiplied by $\Delta t$, respectively.
 The corresponding Cauchy-Green strain tensor (\ref{eq:deformation_tensor}) takes the form
\begin{eqnarray}
{\boldsymbol \Delta} \approx  
   \begin{pmatrix}
    \mathcal{A}&\mathcal{D}&\mathcal{E}\\
   \mathcal{D}& \mathcal{B}& \mathcal{F}\\
    \mathcal{E}&\mathcal{F} &\mathcal{C}
  \end{pmatrix}~,
  \label{eq:general_deformation_tensor}
  \end{eqnarray}
 where, 
 \begin{equation}
 \begin{array}{ll}
 \mathcal{A} = \Gamma_1^2 + \Gamma_3^2 + \omega_1^2~; & \mathcal{D} = \Gamma_1\Gamma_2 + \Gamma_3\Gamma_4 + \omega_1\omega_2~; \\
\mathcal{B} = \Gamma_2^2 + \Gamma_4^2 + \omega_2^2 ~;& \mathcal{E} = \Gamma_1\Lambda_1 + \Gamma_3\Lambda_2 + \omega_1; \\
\mathcal{C} = \Lambda_1^2 + \Lambda_2^2 + 1~; &\mathcal{F} = \Gamma_2\Lambda_1 + \Gamma_4\Lambda_2 + \omega_2 ~;
 \end{array}
 \label{eq:full_eq_set}
 \end{equation}
The characteristic equation of the tensor ${\boldsymbol \Delta}$ in (\ref{eq:general_deformation_tensor}) is 
\begin{eqnarray}
& & \left(\mathcal{A} - \sigma \right)\left[(\mathcal{B}-\sigma) (\mathcal{C}-\sigma)-\mathcal{G}^2\right] \nonumber \\
&-& \mathcal{D}\left[\mathcal{D}(\mathcal{C}-\sigma) - \mathcal{FE} \right] \nonumber \\
&+& \mathcal{E}\left[\mathcal{DF} - \mathcal{E}(\mathcal{B} - \sigma) \right] = 0~, 
\label{eq:characteristic_equation}
\end{eqnarray}
where $\sigma_i$ are the sought eigenvalues. In what follows, different approximations of the parameters in (\ref{eq:full_eq_set}) are made that lead to the FTLE realizations made earlier in (\ref{eq:approx1} - \ref{eq:approx4}). For all approximations, except for approx4, we assume $\Gamma_i=\Gamma,~\Lambda_i=\Lambda,~\omega_i= \omega$.
\begin{itemize}
\item Assuming that  $\omega = 0~, \Lambda \neq  0$, yields approx1. The solutions of the characteristic equation (\ref{eq:characteristic_equation}) are $[0, 4\Gamma^2, 2\Lambda^2+1]$.

\item Assuming that  $\omega \neq 0~, \Lambda =  0$, yields approx2. The solutions of the characteristic equation are thus $[0, 1, 4\Gamma^2 +2\omega] $. 

\item Assuming  $\omega=0, ~\Lambda=0$, yields approx3. The solutions of the characteristic equation are  $[0, 1, 4\Gamma^2]$.

\item Finally, assuming $\Gamma_1=\Gamma_4=1~, \Gamma_3 = \Gamma_2=\omega =0~,$ yields approx4. The solutions of the characteristic equation are $[1, 1, 2\Lambda^2+1]$.
\end{itemize}
In geophysical flows, $\Lambda >> \Gamma$, so that the maximum eingenvalue of approx1 and approx4 is the same and corresponds to $2\Lambda^2+1$. Since $2 \omega^2 \ll 4 \Gamma^2$, approx2 and approx3 also yield he same maximum eigenvalue, that is  $4\Gamma^2$. 
This explains why the numerically computed values of FTLEs are similar for approx1 and approx4 (hereafter called $\lambda_\text{3d}$) and for approx2 and approx 3 (hereafter called $\lambda_\text{2d}$), as visible from Fig. \ref{fig:convergence_of_ftles}d. 
In summary, 
\begin{eqnarray}
\lambda_\text{3d} &\sim& \frac{1}{2\tau}\log \left( 2\Lambda^2 + 1 \right) ~,
\label{eqn:lambdas_3d}\\
\lambda_\text{2d} &\sim& \frac{1}{2\tau}\log \left( 4\Gamma^2 \right) ~.
\label{eqn:lambdas_2d}
\end{eqnarray}
A comparison of the magnitudes of the $\lambda_\text{3d}$ and $\lambda_\text{2d}$ yields
\begin{equation}
\frac{\lambda_\text{3d}}{\lambda_\text{2d}} \sim \log _{4\Gamma^2} \left( 2\Lambda^2 + 1\right),
\label{eqn:lambda_ratio}
\end{equation}
so that $\lambda_\text{3d} \ge \lambda_\text{2d}$ if $2\Lambda^2 + 1 \ge 4\Gamma^2$. The vertical profiles of the horizontally averaged $2\Lambda^2 + 1$ and $4\Gamma^2$ are shown in Fig. \ref{fig:figure4b}, which shows that indeed $2\Lambda^2 + 1 \ge 4\Gamma^2$ at all depths, from which $\lambda_\text{3d} \ge \lambda_\text{2d}$ holds. 

Substituting (\ref{eq:shear}) in (\ref{eqn:lambdas_3d}) leads to,
\begin{equation}
 \qquad \lambda_\text{3d} \sim \frac{1}{2\tau}\text{log}\left[2 \left( \frac{g \Delta t}{ f \rho_{s} \Delta x_i } \right)^2\left(\Delta\rho\right)^2 + 1 \right]~.
 \label{eq:theoretical}
 \end{equation}
Equation (\ref{eq:theoretical}) gives a scaling law between the local FTLEs and the nonlocal density contrast used to initialize the ML front.  

It should be noted that the scaling relation here proposed can be reinterpreted as a relationship between the slope of tracer filaments and $\Delta \rho$. Considering a tracer filament with concentration $C$, the aspect ratio between the horizontal and vertical scales of a tracer filament under the action of horizontal strain and vertical shear, for long time scales yields \citep{haynes_1997,haynes_2001}
\begin{equation}
 \frac{ \partial C / \partial z }{  \nabla_\text{h} C } \sim \frac{\Lambda}{\Gamma} ~,
 \label{eq:slope1}
 \end{equation}
 where $ \nabla_\text{h} = {\bf i}\partial/\partial x + {\bf j}\partial/\partial y~$.
 The same result was found by \citet{smith_2009} only assuming a forward potential enstrophy cascade. In this case,
 \begin{equation}
 \frac{ \partial C / \partial z }{  \nabla_\text{h} C } \sim \frac{N}{f} ~,
 \label{eq:slope1b}
 \end{equation} 
 holds \citep{charney_1971}, as observed in high resolution quasi geostrophic simulations and confirmed from observations of passive tracer dispersion in the North Atlantic \citep{smith_2009}. In our case,
 \begin{equation}
 \frac{ \partial C / \partial z }{  \nabla_\text{h} C } \sim \frac{\Delta x_i}{\Delta z} \sim \left( \frac{g }{ f^2 \rho_{s} \Delta x_i } \right) \Delta \rho ~,
 \label{eq:slope1c}
 \end{equation} 
that can be reduced to (\ref{eq:slope1b}) assuming, without loss of generality, a filament aligned in the zonal direction and using the relationship, valid for the ML \citep{tandon_1994, tandon_1995,young_1994}
\begin{equation}
\left( \frac{b_y}{f} \right)^2 \sim b_z = N^2~.
 \label{eq:slope2}
 \end{equation} 

\begin{figure*}[htb]
\centering
\begin{tabular}{@{}c@{}}                                                   
\noindent\includegraphics[height=0.5\textheight,width=0.48\textwidth]{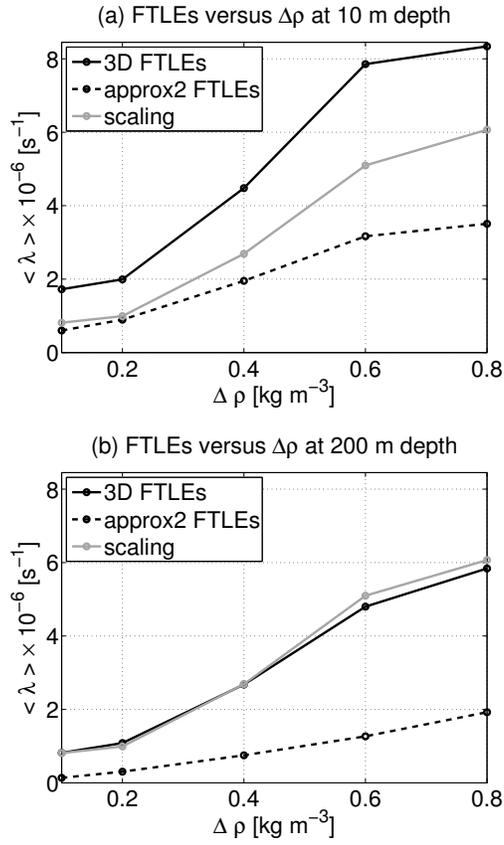}
\end{tabular}
 \caption{Area average of FTLEs versus $\Delta\rho$ for 3D (continuous black line) and approx2 (dashed black lines) FTLEs at the depth of (a) 10 m and (b) 200 m. In gray, the same quantity is shown as derived from the scaling law (\ref{eq:theoretical}).}
  \label{fig:delta_rho}
\end{figure*}

The domain integrated values of the FTLEs $\lambda$ as a function of $\Delta\rho$ shows that, in the ML, the scaling law lies between the 3D and approx2 FTLEs (Fig. \ref{fig:delta_rho}a), while it converges to the values of the 3D FTLEs in the pycnocline (Fig. \ref{fig:delta_rho}b). The large discrepancy between the prediction of the scaling law and the numerical 3D FTLEs in the surface layers is explained by the fact that in the ML, the particle trajectories are dominated by ageostrophic velocities, which are not captured by thermal wind balance. The discrepancy is larger for large values of $\Delta\rho$, corresponding to MLIs with higher values of Rossby numbers. In the pycnocline instead, the ageostrophic component of the flow is weak yielding a  convergence of the scaling law to the 3D FTLEs. 

\subsection{FTLEs Statistics}
Ridges emerging 
from backward FTLEs represent 
regions
to which fluid parcels converge and which are advected by the flow.
It is then possible to consider the backward FTLEs as proxies for a conservative passive tracer, with the FTLE values corresponding to the tracer concentration. 
Under this assumption it is interesting to look at the backward FTLEs statistics, namely the PDFs and the wavenumber spectra, in order to characterize the behavior of the FTLEs. In particular, current parameterizations of passive tracer dispersion by mixed layer instabilities assume the validity of downgradient diffusive schemes \citep{fox-kemper_2008a}, which in turn would imply a Gaussian distribution for the passive tracer, with a Eulerian power spectra following a power law $k^{-1}$, where $k$ is the horizontal wavenumber. While a Gaussian distribution is not expected to hold for the backward FTLEs, it is interesting to calculate their statistics and to compare them between the 3D and 2D case, in order to establish the role of 3D stirring in the distribution of passive tracers.
All the quantities are calculated for the region of the domain where the instabilities are well developed, shown as  the region enclosed by black lines in Fig. \ref{fig:FTLEs}. 

\subsubsection{Probability Distribution Functions}
\label{subsec:PDFs}
\begin{figure*}[htb]
\centering
\begin{tabular}{@{}c@{}}                                                   
\noindent\includegraphics[width=0.45\textwidth]{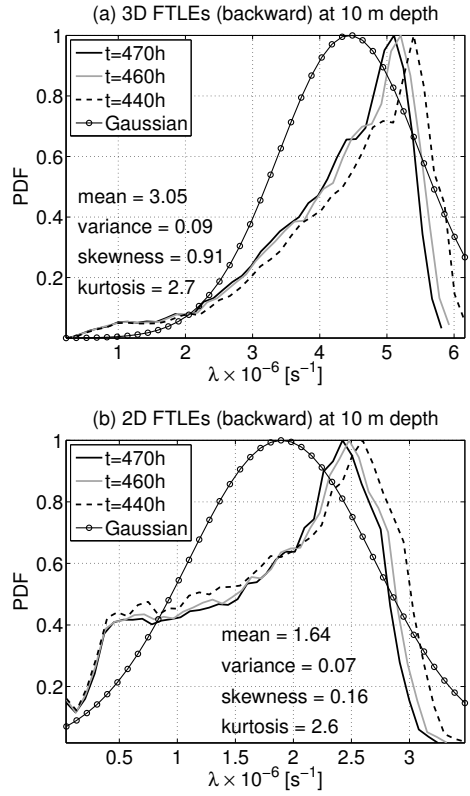}
\end{tabular}
 \caption{PDFs of (a) 3D and (b) 2D FTLEs calculated with backward in time integration at 10~m depth. The FTLEs are calculated using particle integration times of 440 hours (dashed lines), 460 hours (full gray lines) and 470 hours (full black lines). Dotted lines represent the Gaussian distributions with the same mean and standard deviation of the PDFs calculated with the particle integration time of 470 hours.}
  \label{fig:pdfs1}
\end{figure*}

The PDFs of the backward FTLEs calculated for different values of $\tau$ show convergence in time, in agreement with the convergence in time of the horizontally averaged FTLEs (Fig. \ref{fig:pdfs1}). The PDFs of the backward 3D FTLEs at 10 m depth, show large deviations from the Gaussian distribution calculated with the same mean and standard deviation, exhibiting positive values of skewness and long tails toward lower FTLE values (Fig. \ref{fig:pdfs1}a). In comparison, the Gaussian distribution would yield a zero value of skewness. The PDFs are also characterized by low values of kurtosis. In comparison, the Gaussian distribution would yield a value of kurtosis of 3. The PDFs of the backward 3D FTLEs (Fig. \ref{fig:pdfs1}b) show a "shouldering" structure \citep{beron_2008}, which is indicative of a mixed phase space structure of the flow with different attractors and in which different regions experience non uniform stirring rates from the instabilities. While the different shoulders are insufficient to qualify the PDFs of 3D FTLEs as multimodal \citep[e.g.,][]{szezech_2005, harle_2007}, they point to the fact that the stirring in the domain is nonhomogeneous. The deviation from the Gaussian distribution at 10 m depth is visible also in the PDFs of the 2D FTLEs, which show non zero values of skewness and relatively flat peaks corresponding to values of kurtosis larger than 3 (Fig. \ref{fig:pdfs1}b).

\begin{figure*}[htb]
\centering
\begin{tabular}{@{}c@{}}          
\noindent\includegraphics[height=0.5\textheight,width=0.49\textwidth]{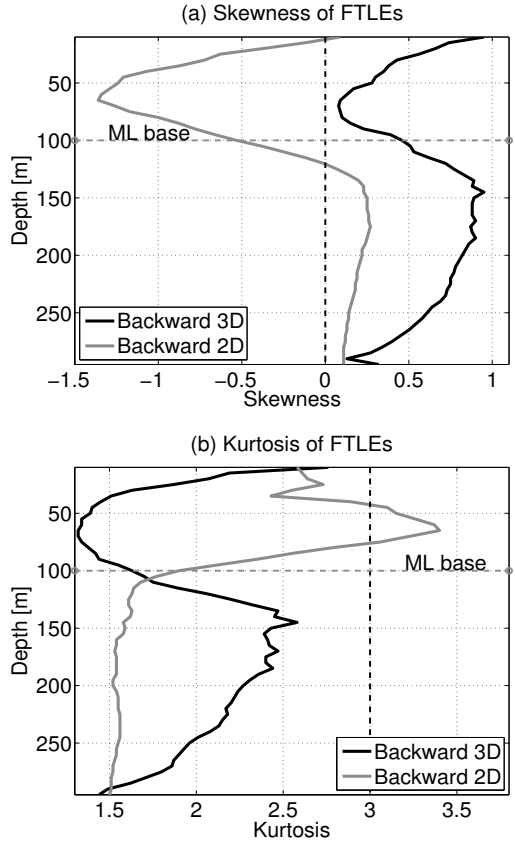}
 \end{tabular}
 \caption{(a) Vertical profiles of the third order moment (skewness) of the PDFs of backward in time 3D FTLEs (black line) and 2D FTLEs (gray line) at day 60 in the reference simulation. The skewness of the Gaussian distribution, equal to  zero is shown as a thin black line. (b) Vertical profiles of the fourth order moment (kurtosis) of the PDFs of the backward in time 3D FTLEs (black line) and 2D FTLEs (gray line). The kurtosis of the Gaussian distribution is 3 (thin black line).}
  \label{fig:moments_profiles}
\end{figure*}

The analysis of the vertical profiles of the skewness and kurtosis of the PDFs of the backward FTLEs, reveals that the distributions of FTLEs are non Gaussian at all depths (Fig. \ref{fig:moments_profiles}). In particular, the skewness of the 3D FTLEs (black lines) shows local maxima at the sea surface and in the pycnocline, and a local minimum within the ML. 
The skewness of the backward in time 2D FTLEs (gray line) shows instead negative values in the ML, increasing to positive values in the interior, with the zero crossing line corresponding to depths just beneath the base of the ML. 

Negative skewed PDFs, as observed in Fig. \ref{fig:pdfs1} at 10 m depth, reveal that most locations in the flow domain experience rates of particle separation greater than the observed average value, with the latter case reflecting a relatively more vigorous stirring influence of the flow. The skewness profiles in Fig. \ref{fig:pdfs1}a show thus that the full 3D stirring leads to higher stirring at all depths than inferred from the 2D approximation. The relatively distinct ridges of the  2D FTLE approximations, particularly in the ML, are reflective of this distribution in which most of the particles experience low stirring rates while a few of them (that lie along ridges) experience higher rates of stirring hence larger values of FTLEs. Non symmetric PDFs, skewed toward low FTLE values have been observed also in previous studies of 2D FTLEs \citep[e.g.,][]{abraham_2002, voth_2002, beron_2009, waugh_2012, harrison_2012}.  

The kurtosis of the 3D FTLEs (Fig. \ref{fig:moments_profiles}b, black line) shows values that are lower than 3 at all depths, corresponding to PDFs that are more peaked than the Gaussian distribution. 
A local minimum is observed at the center of the ML and local maxima are observed at the sea surface and within the pycnocline. 
The kurtosis of the backward in time 2D FTLEs (gray line) shows instead a very different distribution, taking values larger than 3 within the ML, but lower values at the sea surface and in the pycnocline. The low values of kurtosis of PDFs imply that the distributions are relatively flat near the mean value, and thus there is no single dominant phase but an intertwining of multiple phases that contribute to the overall particle separation. In contrast, PDFs with higher values of kurtosis would imply the existence of a dominant phase in a pool of other relatively weaker ones. 

\subsubsection{FTLEs Spectra}
\label{subsec:spectra}
Considering the backward FTLEs as a passive tracer, it is interesting to look at the slopes of the tracer variance, in order to characterize if they show a local or nonlocal behavior. In particular, considering a Eulerian wave number spectra of kinetic energy $E(k) \sim k^{- \alpha}$ and the corresponding tracer spectra $T(k)$, local dynamics are characterized by $1 \le \alpha < 3$, for which the tracer spectra shows a $T(k) \sim k^{\frac{\alpha -3}{2} - 1}$ dependence \citep[e.g.,][]{bennett_1984}. In this regime, the dispersion of particles is dominated by the action of instabilities with size comparable to the separation of the particles. The particular case $T(k) \sim k^{- 2}$ is characteristic of frontal dynamics. For nonlocal dynamics, $\alpha \ge 3$ and $T(k) \sim k^{- 1}$.

The wavenumber spectra are calculated in the zonal direction, i.e. along lines of constant latitude, and then averaged. As for the PDFs, the calculation is performed only in the region occupied by the MLIs, shown between black lines in Fig. \ref{fig:FTLEs}.
  
\begin{figure*}[htb]
\centering
\begin{tabular}{@{}c@{}}      
\noindent\includegraphics[height=0.7\textheight,width=0.8\textwidth]{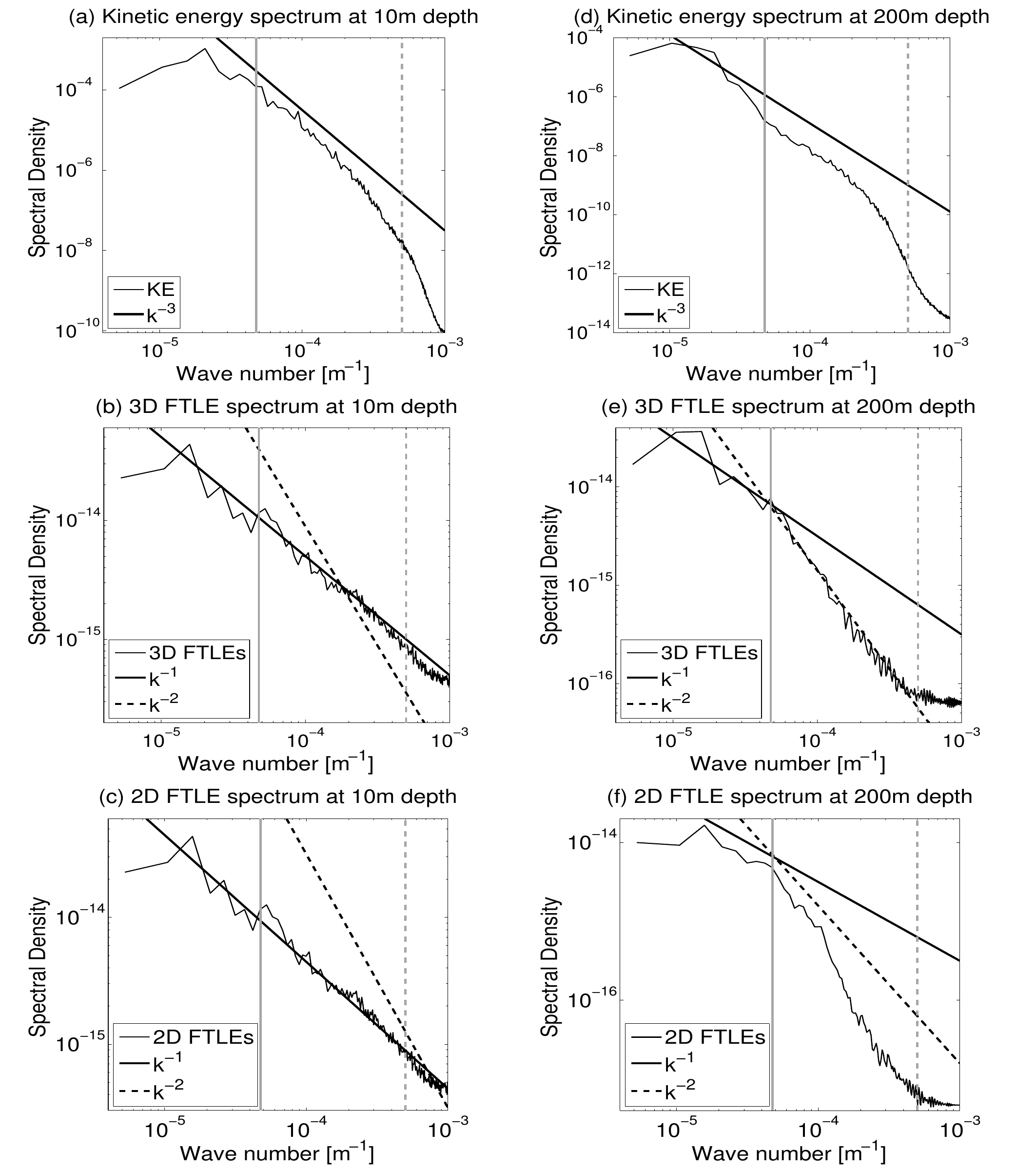}
\end{tabular}
 \caption{ Spectra of (a,d) Kinetic energy, (b,e) backward in time 3D FTLEs and (c,f{)} backward in time 2D FTLEs at 10~m and 200~m depth respectively. The value of the first baroclinic deformation radius (R$_\text{d}$)  in the reference simulation is $\sim$2.06 km (broken gray lines) in the ML and $\sim$21 km (continuous gray lines) in the pycnocline. }
 \label{fig:spectra}
\end{figure*}

In the ML, the kinetic energy (KE) spectra shows slopes of $\alpha = 3$ at scales smaller than the first baroclinic deformation radius, and much steeper slopes at submesoscale, which are thus dominated by dissipation (Fig. \ref{fig:spectra}a). Both the 3D and 2D backward FTLEs spectra show a $- 1$ slope at all scales (Fig. \ref{fig:spectra}b,c), which is in agreement with the slope of the KE spectra and which is a signature of local dispersion created by the mesoscale instabilities. Slopes at smaller scales should instead be interpreted carefully, as at these scales the finite resolution of the model and the numerical dissipation prevent the possible formation of an inertial range. Notice that the 3D and 2D FTLEs spectra display the same pattern of peaks, as a direct consequence of the fact that 3D and 2D FTLEs have ridges in the same locations.

In the pycnocline, the kinetic energy spectrum at scales below the first baroclinic deformation radius shows an inertial range with slope of $\alpha = 3$, or steeper (Fig. \ref{fig:spectra}d). Analysis of the spectra for the backward in time 3D FTLE field reveals however slopes of $\sim $ -2 at 200 m depth (Fig. \ref{fig:spectra}e). The 2D FTLE field at 200 m depth reveals also a $\sim$ -2 slope at scales smaller than the first baroclinic deformation radius, until $\sim$ 10 km, and steeper slopes at smaller scales (Fig. \ref{fig:spectra}f). The spectra slopes of -2 correspond to frontal structures and are in agreement with results from observations from different basins of the World Ocean which show similar slope \citep[e.g.,][]{ferrari_2000,cole_2010,cole_2012,callies_2013,kunze_2015} or even less steep \citep{klymak_2015} both at the surface and in the ocean interior. 
Spectra slopes of -2 were found also from high resolution numerical simulations of the California Current System \citep{capet_2008a}. The spectra suggest that the passive tracer, here characterized from backward in time FTLEs, retains a -2 slope, characteristic of frontal structures \citep{Boyd92}, also at depth, in agreement with the observation that MLIs are able to penetrate in the underlying pycnocline, where they are responsible for horizontal mixing, as observed in numerical simulations by \citet{badin_2011} and in the analytical and semi-analytical solution of \citet{badin13} and \citet{ragone_2015}. It should be noted that this interpretation is challenged by the observations of kinetic energy spectra by \citet{callies_2015}, which suggest instead the predominance of balanced dynamics. \citet{callies_2015} do not, however, examine tracer spectra. Satisfactory scientific explanations on what gives rise to the -2 slope for tracer spectra in the interior are still missing.

The $-1$ slope in the wavenumber spectra at 10 m depth is comparable with the results by \citet{beron_2009}, which found the same slope, representative of local diffusion, at the sea surface. The transition between $-1$ slope close to the sea surface to $-2$ slope at depth can be explained considering that close to the sea surface the flow is more energetic and is responsible for a stronger entanglement of the FTLEs, which results in a larger variance of FTLEs at smaller scales. At depth, FTLEs are less entangled and the spectra displays a smaller variance at small scales. 

It should be noted that the comparison between the results of this study and the results found from observations or from numerical simulations with realistic geometry and forcing is however only of qualitative nature, due to the lack of forcing in the setting here considered.

\subsection{Two-dimensional Lagrangian Coherent Structures}
The chaotic stirring acting on the passive tracer and described in the previous sections is determined by the skeleton of the turbulence underlying the flow. In order to characterize this skeleton of the turbulence, we proceed in calculating the LCSs of the flow under consideration.

\citet{shadden_2005} and \citet{lekien_2007} derive a mathematical framework in which LCSs are extracted as second derivative ridges (or trenches, see, e.g., \citet{beron_2010}) of FTLE fields. However, recently it has been shown that second derivative ridges of FTLE fields predict existence of LCSs in locations where they actually do not exist and fail to yield LCSs in locations where they are known to exist \citep[e.g.,][]{haller_2011, farazmand_2012}. Further studies have claimed that the argument of using second derivative ridges as LCSs is too simplistic and cannot be used for generic flows (e.g., \citet{norgard_2012} and \citet{peikert_2013} for a counter argument). The forementioned shortcomings of extraction of LCSs from FTLE fields have however been addressed in recent studies by defining  LCSs as explicitly parameterized curves derived from invariants of the deformation field  \citep[e.g.,][]{haller_2011, olascoaga_2012, farazmand_2012, beron-vera_2013,blazevski_2014}. The variational theory of LCS extraction specifically targets LCSs as material curves advected by the flow map and also offers the option of obtaining both hyperbolic and elliptic type LCSs as opposed to the FTLE ridge definition, which emphasizes LCSs of hyperbolic type \citep[see ][for a review]{haller_2015}.
 
The variational theory of LCSs provides the necessary and sufficient conditions for the existence of LCSs in terms of the invariants of the Cauchy-Green deformation tensor, and in an objective (i.e frame independent) way \citep[e.g.,][]{haller_2011,farazmand_2012}. Consider the right Cauchy-Green strain tensor
(\ref{eq:deformation_tensor}). In two dimensions, the eigenvalues $\lambda_i$ and eigenvectors  $\xi_i$ of ${\boldsymbol \Delta}\left({\bf x}(t_1), t_1,t_2 \right)$ satisfy the relations,  
\begin{equation}
{\boldsymbol \Delta} \mathbf {\xi}_i = \lambda_i\mathbf{\xi}_i ~,
\label{eq:eigenvalue_problem}
\end{equation}
and
\begin{equation}
\mathbf{\xi}_2 = \Omega \mathbf {\xi}_1~,
\label{eq:eigenvalue_problem2}
\end{equation}
where $0 < \lambda_1 < \lambda_2$, $i=1,2$, and 
\begin{equation}
\Omega=  \begin{pmatrix}
   0&-1\\
   1& 0\\
  \end{pmatrix}~.
\label{eq:eigenvalue_problem_matrix}
\end{equation}
Elliptic LCSs such as vortex boundaries are sought as closed material curves that persist in the flow over the entire integration interval $[t_1,t_2]$ \citep[e.g.,][]{haller_2012,haller_2013}, and have been found to be closed, stationary curves of the averaged tangential stretching functional that coincides with the null geodesics of the Lorentzian metric
\begin{equation}
g_\lambda(u,\upsilon) = \left< u,E_\lambda \upsilon\right>~, 
\label{eq:lorentzian_metric}
\end{equation}   
where $\lambda > 0$ and 
\begin{eqnarray}
E_\lambda({\mathbf x} (t_1)) = \dfrac{1}{2}\left({ \boldsymbol \Delta}(\mathbf{x}(t_1),t_1,t_2) -\lambda^2 I\right)~,
\label{eq:generalized_metric}
\end{eqnarray}
is the generalised Green-Lagrange strain tensor, that measures the deviation of an infinitesimal deformation from a uniform spherical expansion by a factor $\lambda$. The null geodesics resulting from (\ref{eq:lorentzian_metric}) are tangent to the set of vectors,
\begin{eqnarray}
\mathbf{\eta}_\lambda^\pm = \sqrt{\left(\dfrac{\lambda_2(\mathbf x (t_1))-\lambda^2}{\lambda_2(\mathbf x (t_1))-\lambda_1(\mathbf x (t_1))} \right)}\mathbf{\xi}_1(\mathbf x (t_1)) \nonumber \\
\pm \sqrt{\left(\dfrac{\lambda^2 - \lambda_1(\mathbf x (t_1))}{\lambda_2(\mathbf x (t_1))-\lambda_1(\mathbf x (t_1))} \right)}\mathbf{\xi}_2(\mathbf x (t_1))~.
\label{eq:geod_vectors}
\end{eqnarray}
The closed curves corresponding to the outermost $\lambda$ are considered to be the Lagrangian vortex boundaries and are found to satisfy the differential equations
\begin{eqnarray}
r^\prime = \mathbf\eta_{\lambda}^\pm ~,
\label{eq:elliptic_LCSs}
\end{eqnarray}
in which $\lambda$ serves as the parameter  \citep[see][for further discussion]{haller_2015}.
 
Hyperbolic LCSs are defined as stationary curves of an averaged shear functional over the interval $[t_1,t_2]$ which coincides with the null geodesics of the Lorentzian metric \citep[e.g.,][]{farazmand_2014} 
\begin{equation}
g(u,\upsilon) = \left< u,F\upsilon\right>~, 
\label{eq:lorenz_metric2}
\end{equation}
with 
\begin{equation}
F({\mathbf x} (t_1)) = \frac{1}{2}\left({ \boldsymbol \Delta}(\mathbf{x} (t_1))\Omega - \Omega { \boldsymbol \Delta}(\mathbf{x}(t_1))\right) ~,
\label{eq:lorenz_metric2a}
\end{equation}
and $\Omega$ defined in (\ref{eq:eigenvalue_problem}). The geodesic problem in (\ref{eq:lorenz_metric2}) yields a set of differential equations 
\begin{eqnarray}
\mathbf r^\prime_1 = \mathbf\xi_1(r{)} ~, ~ \mathbf r^\prime_2 = \mathbf\xi_2(r{)} ~,
\label{hyperbolic_LCS_eqns}
\end{eqnarray}
from which Repelling and Attracting LCSs are respectively computed as explicitly parameterized curves, with the parameter $r$ being the arc length along the LCS. In the current study, we compute hyperbolic and elliptic LCSs along two dimensional horizontal surfaces implemented with the  LCS Tool - a geodesic LCS detection software for two dimensional unsteady flows \citep{onu_2015}. The integration of stretch and strain line LCSs in (\ref{hyperbolic_LCS_eqns}) is terminated when the arclength parameter $r\ge50$ km in order to ensure a good resolution of the emerging structures. Due to limitations in the computational resources, the LCSs are calculated using the velocity field with 3 hours output, i.e. with a much coarser time resolution than the previous computation, which instead made use of a 15 minutes output.

\begin{figure*}[htb]
\centering
\begin{tabular}{@{}c@{}}      
\includegraphics[height=1.2\textwidth,width=0.5\textwidth]{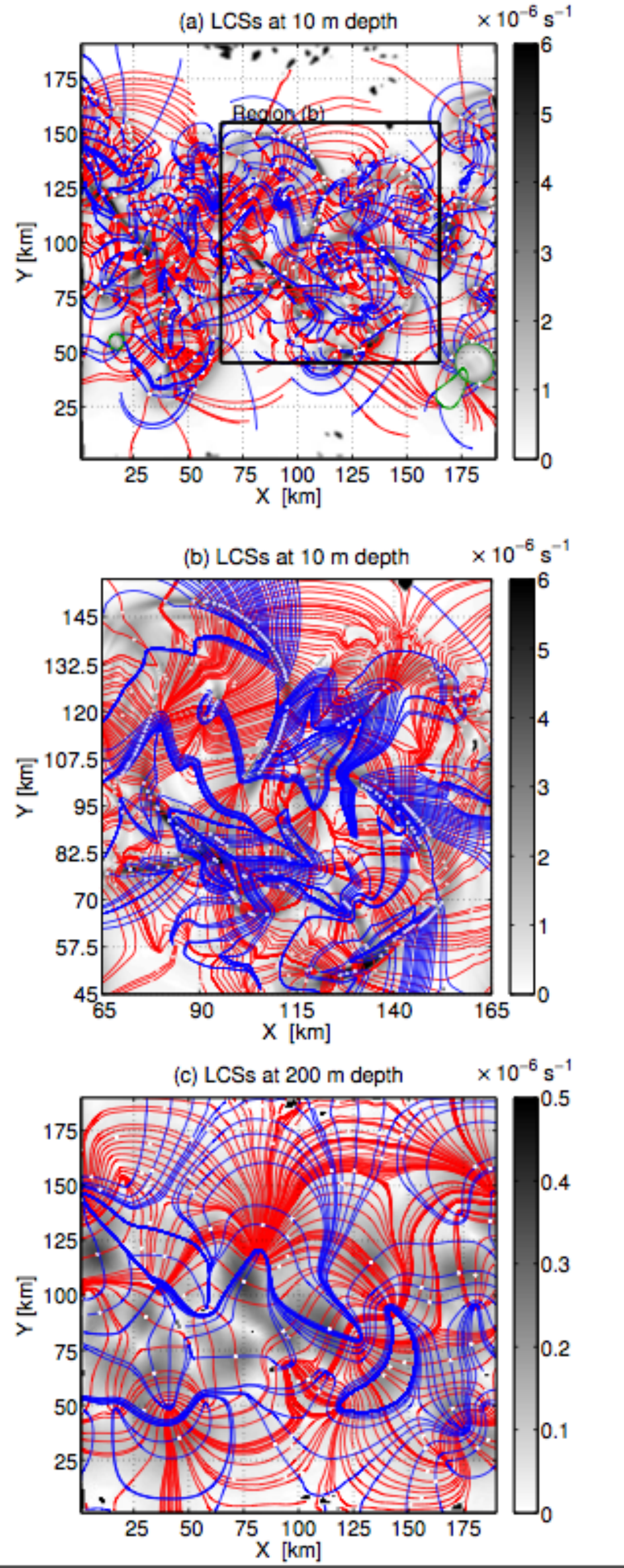}
\end{tabular}
 \caption{ (a) Repelling (red), Attracting (blue) and Elliptic (green) LCSs computed from day 60 to day 80 for the reference run. 2D FTLEs computed for the same period are shown in the background as gray shades. (b) 2D FTLEs and geodesic LCSs in the region demarcated in a black square in panel (a) are computed at double resolution. (c{)} 2D FTLEs and geodesic LCSs at 200 m depth. }
 \label{fig:LCSs}
\end{figure*}

The results for the extraction of the LCSs at 10 m are shown in Fig. \ref{fig:LCSs}a and, in doubled resolution for the region demarcated between the black lines in Fig. \ref{fig:LCSs}a, in Fig. \ref{fig:LCSs}b. Red, blue and green lines indicate respectively Repelling, Attracting and Elliptic LCSs. The FTLEs field is indicated with gray shades. Notice that, due to the different time resolution of the velocity field, the FTLEs field appears smoother than in Fig. \ref{fig:FTLEs}. Frontal structures are observed to be delineated by a complex combination of Repelling and Attracting LCSs, from which a dense network of LCSs spreads over the surrounding regions. The frontal region is also characterized by a web of heteroclinic connections, which form the skeleton of the chaotic flow. As noted from previous studies, the relation between ridges of the FTLE field and the LCSs computed from the variational theory is not one-to-one, although ridges of FTLEs may indicate a nearby LCS \citep{haller_2011, beron-vera_2013}. While ridges of the FTLE field capture most of the important flow features  (especially when computed at high resolution (Fig. \ref{fig:LCSs}b)), it is important to note that the parameterized LCSs offer a more complex structure that cannot be deduced from the ridges of the FTLE field. It should also be noted that the hyperbolic LCSs are dependent on the spatial resolution, with a more convoluted and intricate network of hyperbolic LCSs emerging at higher resolution, with a higher correlation between the FTLE ridges and the hyperbolic LCSs emerging. 

Elliptic LCSs that delineate vortex boundaries, obtained for $\lambda=1$, are represented as closed green curves. Repelling and Attracting LCSs have been truncated so as to start from the boundaries of the elliptic LCSs. The analysis of the Elliptic LCSs confirms the presence of the dipolar structure detaching from the ML front in the lower left corner of the domain, and of another Elliptic structure detached from the frontal region in the lower right side of the domain.

Also in the pycnocline, a complex web of Repelling and Attracting LCSs emerges from the flow. Several regions are observed to be "spreading centres'' of Repelling and Attracting LCSs. Future work will have to determine if these centres evolve into isolated vortices as the flow evolves. The tendency of the geodesically extracted LCSs to predict and reveal flow features and dynamics that are not observed from FTLE fields, allows for a deeper understanding of the Lagrangian skeleton of turbulence \citep{mathur_2007,peacock_2013}. This Lagrangian skeleton leads to the formation of ordered patterns in the flow, and its understanding requires more than the identification of curves of maximal fluid trajectory separation.  

\section{Summary and discussion}
\label{sec:conclusions}

In this study, the 3D FTLEs of ML instabilities have been characterized. Results show that the structure and size of the 3D FTLEs are determined predominantly by the vertical shear of horizontal velocities. 3D FTLE fields exhibit a complex distribution in which high rates of particle separation are not just confined to regions along filaments and vortex boundaries, but are also found in the regions surrounding these high activity features. Regions that are rather quiescent, as observed from Eulerian fields, reveal a complex structure of FTLEs, confirming findings of previous studies which show that a regular flow pattern can yield chaotic particle trajectories \citep[e.g.,][]{aref_1984, ottino_1990, aref_2002, wiggins_2005}. The complexity of the 3D FTLEs field resembles the multifractal distribution of FTLEs found from observations of chaotic stirring by \citet{abraham_2002}. Further, the vertical shear is found to sustain high rates of particle separation in the domain interior. As a consequence, 3D FTLEs decrease slower with depth than 2D FTLEs, which are instead found to be surface intensified and to decrease quickly in magnitude in the pycnocline. It should be noted that 3D and 2D FTLEs display the same spatial distribution of ridges.

The dominating role of vertical shear in the magnitude of the FTLEs is a direct consequence of the nature of MLIs, which is characterized by a stratified and rotating flow in a quasi-balanced state and in which vertical velocities, although larger than their corresponding mesoscale instabilities, is still approximately three orders of magnitude smaller than the horizontal velocities. Analysis of other oceanic flows in which vertical velocities might play an important role, such as coastal upwelling regions, in which vertical velocities are one order of magnitude smaller than the horizontal velocities, reveals a still dominating effect of vertical shear \citep{bettencourt_2012}. It would be interesting to extend the analysis here proposed to other kind of flows, such as idealized flows \citep[e.g.,][]{pratt_2013,rypina_2015} Langmuir turbulence \citep[e.g.,][]{VanRoekel_2012}, in which vertical velocities are comparable to the horizontal velocities and the emerging turbulence is no longer quasi two dimensional.

The observation that 3D FTLEs are dominated by vertical shear allows to determine a scaling relation between the amplitude of the FTLEs and the initial density contrast of the ML front.  While this relationship well agrees with the values of the 3D FTLEs in the interior of the domain, in the ML it shows a deviation from the simulations, which can be attributed to the presence of ageostrophic ML instabilities. 

Backward in time FTLEs can be considered as proxies to a conservative passive tracer, with the FTLE values corresponding to the tracer concentration. Under this assumption it is possible to compare the FTLEs statistics with the statistics expected from passive tracers. PDFs of both 3D and 2D FTLEs are found to be non Gaussian at all depths exhibiting non zero values of skewness and relatively low values of kurtosis. 3D FTLES are skewed toward higher FTLE values with long tails toward low values of FTLEs, while PDFs of 2D FTLEs are instead skewed toward low values of FTLEs with long tails toward higher values of FTLEs. Wavenumber spectra show a slope of -2 in the pycnocline, corresponding to frontal structures and in agreement with results from observations made in various basins of the world ocean, reporting similar spectra slopes for tracers both in the ML and inside the pycnocline\citep[e.g.,][]{ferrari_2000,cole_2010,cole_2012,callies_2013,kunze_2015,klymak_2015}.
The lack of Gaussianity and the slopes of the spectra confirms the observation that the FTLEs possess elongated frontal shapes. Using the 
backward in time FTLEs as proxies for passive tracers, the lack of Gaussianity poses a constraint for the use of diffusive parametrizations, which constrain the stirring effect of MLIs within the ML \citep{fox-kemper_2008a}.

Finally, the observed complex structures of LCSs associated to MLIs can be important for the characterization of mixing and the transfer of nutrients and other passive tracers in the ocean surface, as well as provide the landscape for the growth of different phytoplankton species \citep{dovidio_et_al10}. 


Future studies will have to study carefully the evolution of the LCSs as the MLIs develop. A correct identification of the Elliptic LCSs will allow to calculate an integrated value for the transfer of properties, such as passive and active tracers, from the frontal region. Dynamically, future studies will have to address the influence of forcing of the ML front; the role of seasonality in changing the baroclinicity as well as the vertical shear of the flow; the effects of the coupling of the ML front with the baroclinicity in the pycnocline; and the comparison with realistic simulations and observations, in which all these additional factors, as well as others such as the noise induced by surface winds and internal waves, might change the results found in this study.



\clearpage

\begin{thebibliography}{90}
\providecommand{\natexlab}[1]{#1}
\providecommand{\url}[1]{\texttt{#1}}
\renewcommand{\UrlFont}{\rmfamily}
\providecommand{\urlprefix}{URL }
\expandafter\ifx\csname urlstyle\endcsname\relax
  \providecommand{\doi}[1]{doi:\discretionary{}{}{}#1}\else
  \providecommand{\doi}{doi:\discretionary{}{}{}\begingroup
  \urlstyle{rm}\Url}\fi
\providecommand{\eprint}[2][]{\url{#2}}

\bibitem[{Abraham and Bowen(2002)Abraham, and Bowen}]{abraham_2002}
Abraham, E.~R., and M.~M. Bowen, 2002: Chaotic stirring by a mesoscale
  surface-ocean flow. \textit{Chaos}, \textbf{12}, 373--381.

\bibitem[{Aref(1984)}]{aref_1984}
Aref, H., 1984: Stirring by {C}haotic {A}dvection. \textit{J. Fluid Mech.},
  \textbf{143}, 1 -- 21.

\bibitem[{Aref(2002)}]{aref_2002}
Aref, H., 2002: The development of chaotic advection. \textit{{P}hys.
  {F}luids}, \textbf{14}, 1315 -- 1325.

\bibitem[{Badin(2013)}]{badin13}
Badin, G., 2013: Surface semi-geostrophic dynamics in the ocean.
  \textit{Geophys. Astrophys. Fluid Dyn.}, \textbf{107}, 526--540.

\bibitem[{Badin(2014)}]{badin_2014}
Badin, G., 2014: On the role of non-uniform stratification and short-wave
  instabilities in three-layer quasi-geostrophic turbulence. \textit{{P}hys.
  {F}luids}, \textbf{26}, 096\,603.

\bibitem[{Badin et~al.(2011)Badin, Tandon,, and Mahadevan}]{badin_2011}
Badin, G., A.~Tandon, and A.~Mahadevan, 2011: Lateral {M}ixing in the
  {P}ycnocline by {B}aroclinic {M}ixed {L}ayer {E}ddies. \textit{J. Phys.
  Oceanogr.}, \textbf{41}, 2080 -- 2101.

\bibitem[{Badin et~al.(2009)Badin, Williams, Holt,, and Fernand}]{badin_2009}
Badin, G., R.~G. Williams, J.~T. Holt, and L.~J. Fernand, 2009: Are mesoscale
  eddies in shelf seas formed by baroclinic instability of tidal fronts?
  \textit{J. {G}eophys. {R}es.}, \textbf{114}, C10\,021.

\bibitem[{Badin et~al.(2013)Badin, Williams, Jing,, and Wu}]{badin_2013}
Badin, G., R.~G. Williams, Z.~Jing, and L.~Wu, 2013: Water-mass transformations
  in the {S}outhern {O}cean diagnosed from observations: contrasting effects of
  air-sea fluxes and diapycnal mixing. \textit{J. Phys. Oceanogr.},
  \textbf{43}, 1472--1484.

\bibitem[{Badin et~al.(2010)Badin, Williams,, and Sharples}]{badin_2010}
Badin, G., R.~G. Williams, and J.~Sharples, 2010: {W}ater-mass transformation
  in the shelf seas. \textit{{J}. {M}ar. {R}es.}, \textbf{68}, 189 -- 214.

\bibitem[{Bennett(1984)}]{bennett_1984}
Bennett, A.~F., 1984: Relative dispersion: {L}ocal and nonlocal dynamics.
  \textit{{J}. {A}tmos. {S}ci.}, \textbf{41}, 1881 -- 1886.

\bibitem[{Beron-Vera and Olascoaga(2009)Beron-Vera, and Olascoaga}]{beron_2009}
Beron-Vera, F.~J., and M.~J. Olascoaga, 2009: An {A}ssessment of the
  {I}mportance of {C}haotic {S}tirring and {T}urbulent {M}ixing on the {W}est
  {F}lorida {S}helf. \textit{J. Phys. Oceanogr.}, \textbf{39}, 1743--1755.

\bibitem[{Beron-Vera et~al.(2010)Beron-Vera, Olascoaga, Brown, Ko{\c c}ak,, and
  Rypina}]{beron_2010}
Beron-Vera, F.~J., M.~J. Olascoaga, M.~G. Brown, H.~Ko{\c c}ak, and I.~I.
  Rypina, 2010: Invariant-tori-like {L}agrangian coherent structures in
  geophysical flows. \textit{Chaos}, \textbf{20}, 017\,514.

\bibitem[{Beron-Vera et~al.(2008)Beron-Vera, Olascoaga,, and Goni}]{beron_2008}
Beron-Vera, F.~J., M.~J. Olascoaga, and G.~J. Goni, 2008: Oceanic mesoscale
  eddies as revealed by {L}agrangian coherent structures. \textit{{G}eophys.
  {R}es. {L}ett.}, \textbf{35}, L12\,603.

\bibitem[{Beron-{V}era et~al.(2013)Beron-{V}era, Wang, Olascoaga, Goni,, and
  Haller}]{beron-vera_2013}
Beron-{V}era, F.~J., Y.~Wang, M.~J. Olascoaga, J.~G. Goni, and G.~Haller, 2013:
  Objective detection of oceanic eddies and the {A}gulhas leakage. \textit{J.
  {P}hys. {O}ceanogr.}, \textbf{43}, 1426 -- 1438.

\bibitem[{Bettencourt et~al.(2012)Bettencourt, Lopez,, and
  Hernandez-Garcia}]{bettencourt_2012}
Bettencourt, J.~H., C.~Lopez, and E.~Hernandez-Garcia, 2012: Oceanic
  three-dimensional {L}agrangian coherent structures: {A} study of a mesoscale
  eddy in the {B}enguela upwelling region. \textit{Ocean Modell.}, \textbf{51},
  73 -- 83.

\bibitem[{Blazevski and Haller(2014)Blazevski, and Haller}]{blazevski_2014}
Blazevski, D., and G.~Haller, 2014: Hyperbolic and elliptic transport barriers
  in three dimensional unsteady flows. \textit{Physica {D}}, \textbf{273}, 46
  -- 62.

\bibitem[{Boccaletti et~al.(2007)Boccaletti, Ferrari,, and
  Fox-Kemper}]{boccaletti_2007}
Boccaletti, G., R.~Ferrari, and B.~Fox-Kemper, 2007: Mixed {L}ayer
  {I}nstabilities and {R}estratification. \textit{J. Phys. Oceanogr.},
  \textbf{37}, 2228 -- 2250.

\bibitem[{Boffetta et~al.(2001)Boffetta, Lacorata, Redaelli,, and
  Vulpiani}]{boffetta_2001}
Boffetta, G., G.~Lacorata, G.~Redaelli, and A.~Vulpiani, 2001: Detecting
  barriers to transport: a review of different techniques. \textit{Physica D},
  \textbf{159}, 58--70.

\bibitem[{Boyd(1992)}]{Boyd92}
Boyd, J., 1992: The energy spectrum of fronts: time evolution and shocks in
  {B}urgers' equation. \textit{J. Atmos. Sci.}, \textbf{49}, 128--139.

\bibitem[{Calil and Richards(2010)Calil, and Richards}]{calil_2010}
Calil, P.~H., and K.~J. Richards, 2010: Transient upwelling hot spots in the
  oligotrophic {N}orth {P}acific. \textit{J. {G}eophys. {R}es.}, \textbf{115},
  C02\,003.

\bibitem[{Callies and Ferrari(2013)Callies, and Ferrari}]{callies_2013}
Callies, J., and R.~Ferrari, 2013: Interpreting {E}nergy and {T}racer {S}pectra
  of {U}pper-{O}cean {T}urbulence in the {S}ubmesoscale {R}ange (1-200 km).
  \textit{J. {P}hys. {O}ceanogr.}, \textbf{43}, 2456 -- 2474.

\bibitem[{Callies et~al.(2015)Callies, Ferrari, Klimak,, and
  Gula}]{callies_2015}
Callies, J., R.~Ferrari, J.~M. Klimak, and J.~Gula, 2015: Seasonality in
  submesoscale turbulence. \textit{Nature Communications}, \textbf{6}, 6862.

\bibitem[{Capet et~al.(2008)Capet, McWilliams, Molemaker,, and
  Shchepetkin}]{capet_2008a}
Capet, X., J.~C. McWilliams, M.~J. Molemaker, and A.~F. Shchepetkin, 2008:
  Mesoscale to {S}ubmesoscale {T}ransition in the {C}alifornia {C}urrent
  {S}ystem. {P}art {I}: {F}low {S}tructure, {E}ddy {F}lux and {O}bservational
  {T}ests. \textit{J. Phys. Oceanogr.}, \textbf{38}, 29 -- 43.

\bibitem[{Charney(1971)}]{charney_1971}
Charney, J., 1971: Geostrophic {T}urbulence. \textit{{J}. {A}tmos. {S}ci.},
  \textbf{28}, 1087 -- 1095.

\bibitem[{Cole and Rudnick(2012)Cole, and Rudnick}]{cole_2012}
Cole, S.~T., and D.~L. Rudnick, 2012: The spatial distribution and annual cycle
  of upper-ocean thermohaline structure. \textit{J. {G}eophys. {R}es.},
  \textbf{117~(C02027)}, \doi{doi:10.1029/2011JC007033}.

\bibitem[{Cole et~al.(2010)Cole, Rudnick,, and Colosi}]{cole_2010}
Cole, S.~T., D.~L. Rudnick, and J.~Colosi, 2010: Seasonal evolution of upper
  ocean horizontal structure and the remnant mixed layer. \textit{J. {G}eophys.
  {R}es.}, \textbf{115~(C04012)}, \doi{doi:10.1029/2009JC005654}.

\bibitem[{d'Ovidio et~al.(2009)d'Ovidio, Isern-Fontanet, Lopez,
  Hernandez-Garcia,, and Garcia-Ladon}]{dovidio_2009}
d'Ovidio, F., J.~Isern-Fontanet, C.~Lopez, E.~Hernandez-Garcia, and
  E.~Garcia-Ladon, 2009: Comparison between {E}ulerian diagnostics and
  finite-size {L}yapunov exponents computed from altimetry in the {A}lgerian
  basin. \textit{Deep-Sea Res. I}, \textbf{56}, 15 -- 31.

\bibitem[{d'Ovidio et~al.(2010)d'Ovidio, Monte, Alvain, Dandonneau,, and
  L\'{e}vy}]{dovidio_et_al10}
d'Ovidio, F., S.~D. Monte, S.~Alvain, Y.~Dandonneau, and M.~L\'{e}vy, 2010:
  Fluid dynamical niches of phytoplankton types. \textit{PNAS}, \textbf{107},
  18\,366 -- 18\,370.

\bibitem[{Farazmand et~al.(2014)Farazmand, Blazevski,, and
  Haller}]{farazmand_2014}
Farazmand, M., D.~Blazevski, and G.~Haller, 2014: Shearless transport barriers
  in unsteady two-dimensional flows and maps. \textit{Physica {D}},
  \textbf{278-279}, 44 -- 57.

\bibitem[{Farazmand and Haller(2012)Farazmand, and Haller}]{farazmand_2012}
Farazmand, M., and G.~Haller, 2012: Computing {L}agrangian {C}oherent
  {S}tructures from their variational theory. \textit{Chaos},
  \textbf{22~(013128)}.

\bibitem[{Ferrari and Rudnick(2000)Ferrari, and Rudnick}]{ferrari_2000}
Ferrari, R., and D.~L. Rudnick, 2000: Thermohaline variability in the upper
  ocean. \textit{J. {G}eophys. {R}es.}, \textbf{105}, 16\,857 -- 16\,883.

\bibitem[{Fox-{K}emper and Ferrari(2008)Fox-{K}emper, and
  Ferrari}]{fox-kemper_2008b}
Fox-{K}emper, B., and R.~Ferrari, 2008: Parameterization of {M}ixed {L}ayer
  {E}ddies. {P}art {II}: {P}rognosis and {I}mpact. \textit{J. Phys. Oceanogr.},
  \textbf{38}, 1166 -- 1179.

\bibitem[{Fox-Kemper et~al.(2008)Fox-Kemper, Ferrari,, and
  Hallberg}]{fox-kemper_2008a}
Fox-Kemper, B., R.~Ferrari, and R.~Hallberg, 2008: Parameterization of {M}ixed
  {L}ayer {E}ddies. {P}art {I}: {T}heory and {D}iagnosis. \textit{J. Phys.
  Oceanogr.}, \textbf{38}, 1145 -- 1165.

\bibitem[{Gula et~al.(2014)Gula, Molemaker,, and McWilliams}]{gula_2014}
Gula, J., M.~J. Molemaker, and J.~C. McWilliams, 2014: {S}ubmesoscale {C}old
  {F}ilaments in the {G}ulf {S}tream. \textit{J. {P}hys. {O}ceanogr.},
  \textbf{44}, 2617 -- 2643.

\bibitem[{Haine and Marshall(1998)Haine, and Marshall}]{haine_1998}
Haine, T. W.~N., and J.~Marshall, 1998: Gravitational, {S}ymmetric and
  {B}aroclinic {I}nstability of the {O}cean {M}ixed {L}ayer. \textit{J. Phys.
  Oceanogr.}, \textbf{28}, 634 -- 658.

\bibitem[{Haller(2000)}]{haller_2000b}
Haller, G., 2000: Finding finite-time invariant manifolds in two-dimensional
  velocity fields. \textit{Chaos}, \textbf{10}, 99--108.

\bibitem[{Haller(2001)}]{haller_2001}
Haller, G., 2001: Distinguished material surfaces and coherent structures in
  three-dimensional fluid flows. \textit{Physica D}, \textbf{149}, 248--277.

\bibitem[{Haller(2002)}]{haller_2002}
Haller, G., 2002: Lagrangian coherent structures from approximate velocity
  data. \textit{Phys. Fluids.}, \textbf{14}, 1851--1861.

\bibitem[{Haller(2011)}]{haller_2011}
Haller, G., 2011: A variational theory of hyperbolic {L}agrangian {C}oherent
  {S}tructures. \textit{Physica {D}}, \textbf{240}, 574 -- 598.

\bibitem[{Haller(2015)}]{haller_2015}
Haller, G., 2015: {L}agrangian {C}oherent {S}tructures. \textit{Ann. {R}ev.
  {F}luid {M}ech.}, \textbf{47}, 137 -- 162.

\bibitem[{Haller and Beron-{V}era(2012)Haller, and Beron-{V}era}]{haller_2012}
Haller, G., and F.~J. Beron-{V}era, 2012: Geodesic theory of transport barriers
  in two dimensional flows. \textit{Physica {D}}, \textbf{241}, 1680 -- 1702.

\bibitem[{Haller and Beron-{V}era(2013)Haller, and Beron-{V}era}]{haller_2013}
Haller, G., and F.~J. Beron-{V}era, 2013: Coherent {L}agrangian vortices: the
  black holes of turbulence. \textit{J. {F}luid {M}ech.},
  \textbf{731~(R4:1-10)}.

\bibitem[{Haller and Yuan(2000)Haller, and Yuan}]{haller_2000a}
Haller, G., and G.~Yuan, 2000: Lagrangian coherent structures and mixing in
  two-dimensional turbulence. \textit{Physica D}, \textbf{147}, 352--370.

\bibitem[{Harle and Feudel(2007)Harle, and Feudel}]{harle_2007}
Harle, M., and U.~Feudel, 2007: Hierachy of islands in conservative systems
  yields multimodal distributions of {FTLE}s. \textit{{C}haos {S}olitons
  {F}ractals}, \textbf{31}, 130 -- 137.

\bibitem[{Harrison and Glatzmaier(2012)Harrison, and
  Glatzmaier}]{harrison_2012}
Harrison, C.~S., and G.~A. Glatzmaier, 2012: {L}agrangian {C}oherent
  {S}tructures in the {C}alifornia {C}urrent {S}ystem - {S}ensitivies and
  {L}imitations. \textit{Geophys. {A}strophys. Fluid. {D}yn.}, \textbf{106},
  22--24.

\bibitem[{Haynes and Anglade(1997)Haynes, and Anglade}]{haynes_1997}
Haynes, P., and J.~Anglade, 1997: The {V}ertical-{S}cale {C}ascade in
  {A}tmospheric {T}racers due to {L}arge-{S}cale {D}ifferential {A}dvection.
  \textit{{J}. {A}tmos. {S}ci.}, \textbf{54}, 1121 -- 1136.

\bibitem[{Haynes(2001)}]{haynes_2001}
Haynes, P.~H., 2001: Vertical shear plus horizontal stretching as a route to
  mixing. \textit{Proceedings. 12$^{th}$ Hawaii {W}inter Workshop (16 - 19
  January)}, University of Hawaii at Manoa.

\bibitem[{Klymak et~al.(2015)Klymak, Crawford, Alford, MacKinnon,, and
  Pinkel}]{klymak_2015}
Klymak, J.~M., W.~Crawford, M.~H. Alford, J.~A. MacKinnon, and R.~Pinkel, 2015:
  Along-isopycnal variability of spice in the {N}orth {P}acific. \textit{J.
  {G}eophys. {R}es. {O}ceans.}, \textbf{120}, 2287 -- 2307.

\bibitem[{Kunze et~al.(2015)Kunze, Klymak, Lien, Ferrari, Lee, Sundermeyer,,
  and Goodman}]{kunze_2015}
Kunze, E., J.~M. Klymak, R.~C. Lien, R.~Ferrari, C.~M. Lee, M.~A. Sundermeyer,
  and L.~Goodman, 2015: Submesoscale {W}ater-{M}ass {S}pectra in the {S}argasso
  {S}ea. \textit{J. {P}hys. {O}ceanogr.}, \textbf{45}, 1325 -- 1338.

\bibitem[{Lapeyre(2002)}]{lapeyre_2002}
Lapeyre, G., 2002: Characterization of finite-time {L}yapunov exponents and
  vectors in two-dimensional turbulence. \textit{Chaos}, \textbf{12}, 688 --
  698.

\bibitem[{Lekien et~al.(2007)Lekien, Shadden,, and Marsden}]{lekien_2007}
Lekien, F., S.~C. Shadden, and J.~E. Marsden, 2007: Lagrangian coherent
  structures in n-dimensional systems. \textit{{J}. {M}ath. {P}hys.},
  \textbf{48}, 065\,404.

\bibitem[{L{\'e}vy et~al.(2001)L{\'e}vy, Klein,, and Treguier}]{levy_2001}
L{\'e}vy, M., P.~Klein, and A.~Treguier, 2001: {I}mpacts of {S}ubmesoscale
  {P}hysics on {P}roduction and {S}ubduction of {P}hytoplankton in an
  {O}ligotrophic {R}egime. \textit{{J}. {M}ar. {R}es.}, \textbf{59}, 535 --
  565.

\bibitem[{Mahadevan(2006)}]{mahadevan_2006b}
Mahadevan, A., 2006: Modelling vertical motion at ocean fronts: Are
  nonhydrostatic effects relevant at submesoscales? \textit{Ocean {M}odell.},
  \textbf{14}, 222 --240.

\bibitem[{Mahadevan and Tandon(2006)Mahadevan, and Tandon}]{mahadevan_2006a}
Mahadevan, A., and A.~Tandon, 2006: An analysis of mechanisms for submesoscale
  vertical motion at ocean fronts. \textit{Ocean {M}odell.}, \textbf{14}, 241
  -- 256.

\bibitem[{Mahadevan et~al.(2010)Mahadevan, Tandon,, and
  Ferrari}]{mahadevan_2010}
Mahadevan, A., A.~Tandon, and R.~Ferrari, 2010: Rapid changes in mixed layer
  stratification driven by submesoscale instabilities and winds. \textit{J.
  {G}eophys. {R}es.}, \textbf{115}, C03\,017.

\bibitem[{Marshall et~al.(1997{\natexlab{a}})Marshall, A.~Adcroft,, and
  Heisey}]{mitgcm_1997b}
Marshall, J., L.~P. A.~Adcroft, C.~Hill, and C.~Heisey, 1997{\natexlab{a}}: A
  finite-volume, incompressible {N}avier {S}tokes model for studies of the
  ocean on parallel computers. \textit{J. {G}eophys. {R}es.}, \textbf{102},
  5753 -- 5766.

\bibitem[{Marshall et~al.(1997{\natexlab{b}})Marshall, Hill, Perelman,, and
  Adcroft}]{mitgcm_1997a}
Marshall, J., C.~N. Hill, L.~Perelman, and A.~Adcroft, 1997{\natexlab{b}}:
  Hydrostatic, quasi - hydrostatic and non - hydrostatic ocean modelling.
  \textit{J. {G}eophys. {R}es.}, \textbf{102}, 5753 -- 5752.

\bibitem[{Mathur et~al.(2007)Mathur, Haller, Peacock, Rupert-{F}elsot,, and
  Swinney}]{mathur_2007}
Mathur, M., G.~Haller, T.~Peacock, J.~E. Rupert-{F}elsot, and H.~L. Swinney,
  2007: Uncovering the {L}agrangian {S}keleton of {T}urbulence. \textit{{P}hys.
  {R}ev. {L}ett.}, \textbf{98}, 144\,502.

\bibitem[{Molemaker and McWilliams(2005)Molemaker, and
  McWilliams}]{molemaker_2005}
Molemaker, M.~J., and J.~C. McWilliams, 2005: {B}aroclinic {I}nstability and
  {L}oss of {B}alance. \textit{J. Phys. Oceanogr.}, \textbf{35}, 1505 -- 1517.

\bibitem[{Norgard and Bremer(2012)Norgard, and Bremer}]{norgard_2012}
Norgard, G., and P.~Bremer, 2012: Second derivative ridges are straight lines
  and the implications for computing {L}agrangian {C}oherent {S}tructures.
  \textit{Physica {D}}, \textbf{241}, 1475 -- 1476.

\bibitem[{Okubo(1970)}]{okubo_1970}
Okubo, A., 1970: Horizontal dispersion of floatable trajectories in the
  vicinity of velocity singularities such as convergences. \textit{{D}eep-{S}ea
  {R}es.}, \textbf{17}, 445 -- 454.

\bibitem[{Olascoaga and Haller(2012)Olascoaga, and Haller}]{olascoaga_2012}
Olascoaga, M.~J., and G.~Haller, 2012: Forecasting sudden changes in
  environmental pollution patterns. \textit{PNAS}, \textbf{109}, 4738 -- 4743.

\bibitem[{Onu et~al.(2015)Onu, Huhn,, and Haller}]{onu_2015}
Onu, K., F.~Huhn, and G.~Haller, 2015: {LCS} {T}ool: {A} computational platform
  for {L}agrangian {C}oherent {S}tructures. \textit{J. {C}omput. {S}ci.},
  \textbf{7}, 26 -- 36.

\bibitem[{Ottino(1990)}]{ottino_1990}
Ottino, J.~M., 1990: {Mixing, Chaotic Advection and Turbulence}. \textit{{A}nn.
  {R}ev. {F}luid {M}ech.}, \textbf{22}, 207 -- 254.

\bibitem[{Peacock and Haller(2013)Peacock, and Haller}]{peacock_2013}
Peacock, T., and G.~Haller, 2013: {L}agrangian coherent structures: {T}he
  hidden skeleton of fluid flows. \textit{{P}hysics {T}oday}, \textbf{66}, 41
  -- 47.

\bibitem[{Peikert et~al.(2013)Peikert, G{\"u}nther,, and
  Weinkauf}]{peikert_2013}
Peikert, R., D.~G{\"u}nther, and T.~Weinkauf, 2013: Comment on "{S}econd
  derivative ridges are straight lines and the implications for computing
  {L}agrangian {C}oherent {S}tructures, {P}hysica {D} 2012.05.006".
  \textit{Physica {D}}, \textbf{242}, 65 -- 66.

\bibitem[{Pratt et~al.(2013)Pratt, Rypina, {\"O}ezg{\"o}kmen, Wang, Childs,,
  and Bebieva}]{pratt_2013}
Pratt, L.~J., I.~I. Rypina, T.~M. {\"O}ezg{\"o}kmen, P.~Wang, H.~Childs, and
  Y.~Bebieva, 2013: Chaotic advection in a steady, three-dimensional,
  {E}kman-driven eddy. \textit{J. {F}luid {M}ech.}, \textbf{738}, 143 -- 183.

\bibitem[{Price(1981)}]{price_1981}
Price, J.~F., 1981: Upper ocean response to a hurricane. \textit{J. {P}hys.
  {O}ceanogr.}, \textbf{11}, 153 -- 175.

\bibitem[{Ragone and Badin(2016)Ragone, and Badin}]{ragone_2015}
Ragone, F., and G.~Badin, 2016: {S}urface semi-geostrophic turbulence:
  freely-evolving dynamics. \textit{J. {F}luid {M}ech.}, \textbf{792}, 740 -- 774.

\bibitem[{Rypina et~al.(2007)Rypina, Brown, Beron-Vera, Ko{\c c}ak, Olascoaga,,
  and Udovydchenkov}]{rypina_2007}
Rypina, I.~I., M.~G. Brown, F.~J. Beron-Vera, H.~Ko{\c c}ak, M.~J. Olascoaga,
  and I.~A. Udovydchenkov, 2007: On the {L}agrangian {D}ynamics of
  {A}tmospheric {Z}onal {J}ets and the {P}ermeability of the {S}tratospheric
  {P}olar {V}ortex. \textit{{J}. {A}tmos. {S}ci.}, \textbf{64}, 3595 -- 3610.

\bibitem[{Rypina et~al.(2010)Rypina, Pratt, Pullen, Levin,, and
  Gordon}]{rypina_2010}
Rypina, I.~I., L.~J. Pratt, J.~Pullen, J.~Levin, and A.~L. Gordon, 2010:
  Chaotic {A}dvection in an {A}rchipelago. \textit{J. Phys. Oceanogr.},
  \textbf{40}, 1988 -- 2006.

\bibitem[{Rypina et~al.(2015)Rypina, Pratt, Wang, Oezgoekmen,, and
  Mezic}]{rypina_2015}
Rypina, I.~I., L.~J. Pratt, P.~Wang, T.~Oezgoekmen, and I.~Mezic, 2015:
  Resonance phenomena in a time-dependent, three-dimensional model of an
  idealized eddy. \textit{CHAOS}, \textbf{25}, 087\,401.

\bibitem[{Shadden et~al.(2005)Shadden, Lekien,, and Marsden}]{shadden_2005}
Shadden, S.~C., F.~Lekien, and J.~E. Marsden, 2005: Definition and properties
  of {L}agrangian {C}oherent structures from finite-time {L}yapunov exponents
  in two-dimensional aperiodic flows. \textit{Physica D}, \textbf{212},
  271--304.

\bibitem[{Shcherbina et~al.(2015)}]{shcherbina_2015}
Shcherbina, A., and Coauthors, 2015: The {L}atmix {S}ummer {C}ampaign:
  {S}ubmesoscale {S}tirring in the {U}pper {O}cean. \textit{Bull. Amer. Meteor.
  Soc.}, \textbf{96}, 1257 -- 1279, \doi{10.1175/BAMS-D-14-00015.1}.

\bibitem[{Smith and Ferrari(2009)Smith, and Ferrari}]{smith_2009}
Smith, K.~S., and R.~Ferrari, 2009: The {P}roduction and {D}issipation of
  {C}ompensated {T}hermohaline {V}ariance by {M}esoscale {S}tirring. \textit{J.
  {P}hys. {O}ceanogr.}, \textbf{39}, 2477 -- 2501.

\bibitem[{Sulman et~al.(2013)Sulman, Huntley, {J}r,, and {J}r}]{sulman_2013}
Sulman, M. H.~M., H.~S. Huntley, B.~L.~L. {J}r, and A.~D.~K. {J}r, 2013:
  Leaving flatland: {D}iagnostics for {L}agrangian coherent structures in
  three-dimensional flows. \textit{Physica D}, \textbf{258}, 77--92.

\bibitem[{Szezech et~al.(2005)Szezech, Lopes,, and Viana}]{szezech_2005}
Szezech, J. D.~J., S.~R. Lopes, and R.~L. Viana, 2005: Finite-time {L}yapunov
  spectrum for chaotic orbits of non-integrable {H}amiltonian systems.
  \textit{Phys {L}ett. A}, \textbf{335}, 394 -- 401.

\bibitem[{Tandon and Garrett(1994)Tandon, and Garrett}]{tandon_1994}
Tandon, A., and C.~Garrett, 1994: {M}ixed {L}ayer {R}estratification due to a
  {H}orizontal {D}ensity {G}radient. \textit{J. Phys. Oceanogr.}, \textbf{24},
  1419 -- 1424.

\bibitem[{Tandon and Garrett(1995)Tandon, and Garrett}]{tandon_1995}
Tandon, A., and C.~Garrett, 1995: Geostrophic {A}djustment and
  {R}estratification of a {M}ixed {L}ayer with {H}orizontal {G}radients above a
  {S}tratified {L}ayer. \textit{J. Phys. Oceanogr.}, \textbf{25}, 2229 -- 2241.

\bibitem[{Taylor and Ferrari(2009)Taylor, and Ferrari}]{taylor_2009}
Taylor, J.~R., and R.~Ferrari, 2009: On the equillibration of a symmetrically
  unstable front via a secondary shear instability. \textit{J. {F}luid
  {M}ech.}, \textbf{622}, 103 -- 113.

\bibitem[{Thomas and Joyce(2010)Thomas, and Joyce}]{Thomas_Joyce_2010}
Thomas, L.~N., and T.~M. Joyce, 2010: Subduction in the northern and southern
  flanks of the {G}ulf {S}tream. \textit{J. Phys. Oceanogr.}, \textbf{40},
  429--438.

\bibitem[{Thomas et~al.(2008)Thomas, Tandon,, and Mahadevan}]{leif_2008}
Thomas, L.~N., A.~Tandon, and A.~Mahadevan, 2008: {S}ubmesoscale {P}rocesses
  and {D}ynamics. \textit{Ocean {M}odell.}, \textbf{177}, 17 --38.

\bibitem[{Thomas et~al.(2013)Thomas, Taylor, Ferrari,, and Joyce}]{Thomas_2013}
Thomas, L.~N., J.~R. Taylor, R.~Ferrari, and T.~M. Joyce, 2013: Symmetric
  instability in the {G}ulf {S}tream. \textit{Deep-Sea Res. II}, \textbf{91},
  96--110.

\bibitem[{Van~Roekel et~al.(2012)Van~Roekel, Fox-Kemper, Sullivan, Hamlington,,
  and Haney}]{VanRoekel_2012}
Van~Roekel, L.~P., B.~Fox-Kemper, P.~P. Sullivan, P.~E. Hamlington, and S.~R.
  Haney, 2012: The form and orientation of langmuir cells for misaligned winds
  and waves. \textit{J. Geophys. Res.}, \textbf{117}, C05\,001.

\bibitem[{Voth et~al.(2002)Voth, Haller,, and Gollub}]{voth_2002}
Voth, G.~A., G.~Haller, and J.~P. Gollub, 2002: Experimental {M}easurements of
  {S}tretching {F}ields in {F}luid {M}ixing. \textit{{P}hys. {R}ev. {L}ett.},
  \textbf{88}, 254\,501.

\bibitem[{Waugh and Abraham(2008)Waugh, and Abraham}]{waugh_2008}
Waugh, D.~W., and E.~R. Abraham, 2008: Stirring in the global surface ocean.
  \textit{{G}eophys. {R}es. {L}ett.}, \textbf{35}, L20\,605.

\bibitem[{Waugh et~al.(2012)Waugh, Keating,, and Chen}]{waugh_2012}
Waugh, D.~W., S.~R. Keating, and M.~Chen, 2012: Diagnosing {O}cean {S}tirring:
  {C}omaparison of {R}elative {D}ispersion and {F}inite-{T}ime {L}yapunov
  {E}xponents. \textit{J. Phys. Oceanogr.}, \textbf{42}, 1173--1185.

\bibitem[{Weiss(1991)}]{weiss_1991}
Weiss, J., 1991: The {D}ynamics of {E}nstrophy transfer in two-dimensional
  hydrodynamics. \textit{Physica D}, \textbf{48}, 273 -- 294.

\bibitem[{Wiggins(2005)}]{wiggins_2005}
Wiggins, S., 2005: {T}he {D}ynamical {S}ystems {A}pproach {T}o {L}agrangian
  {T}ransport in {O}ceanic {F}lows. \textit{Ann. {R}ev. {F}luid {M}ech.},
  \textbf{37}, 295 -- 328.

\bibitem[{Young(1994)}]{young_1994}
Young, W.~R., 1994: The subinertial mixed layer approximation. \textit{J.
  {P}hys. {O}ceanogr.}, \textbf{24}, 1812 -- 1826.

\end{thebibliography}

\begin{table*}[ht!]
\begin{center}
\caption{Table of model parameters.}
\begin{tabular}{l c c }
\hline \hline 
Parameter              &    Symbol      &    Value  \\  
\hline 
Coriolis parameter   & $f$                  &  $1.0284 \times 10^{-4}$  s$^{-1}$ \\ 
Horizontal, meridional lengths of the channel & L$_x$, L$_y$       & 192                   km \\
Depth of the channel                 & $H_{tot}$               & 300                    m \\
Mixed layer depth                      & $H_{ML}$ & 100                 m \\
Spatial resolution                          & (dx, dy, dz)   & (500, 500, 5)      m \\
Lateral biharmonic viscosity      & $\nu_\text{H}$    &  $2 \times 10^{5}$  m$^4$ s$^{-1}$ \\
Vertical eddy viscosity                     & $\nu_\text{v}$    & $10^{-4}$   m$^2$ s$^{-1}$  \\
Lateral biharmonic diffusivity of heat, salt &  $\text{K}_\text{T}$, $\text{K}_\text{S}$  &$10^{2}$     m$^4$ s$^{-1}$ \\
Vertical diffusivity of temperature, salt     & $\text{K}_\text{Tz}$, $\text{K}_\text{Sz}$   & $10^{-5}$   m$^2$ s$^{-1}$  \\
\hline \hline 
\end{tabular}
\label{tab:model_parameters}
\end{center}
\end{table*}

\clearpage 
\begin{table*}[htb!]
\begin{center}
\caption{Numerical experiments conducted and the time windows during which particle trajectories are computed.}
\begin{tabular}{c c  c c c }
\hline \hline 
$\Delta\rho$      & Time window (days)  &     Deformation radii [km]\\
&                          $[t_1,t_2]$                                            & $\quad$ ~ML $\quad$ ~Pycnocline\\
\hline 
0.1                     &  285  - 330                & 1.00 $\qquad$ ~21.75\\     
0.2                     &  165  - 210                & 1.45 $\qquad$ ~21.70\\    
\hline
0.4                   &   60  - 80                    & 2.06 ~$\qquad$ 21.55 \\
(reference simulation)  &                    &                                                             \\  
\hline
0.6                     &   45  - 60                   & 2.16 $\qquad$ 21.35\\     
0.8                     &   45  - 60                   & 3.91 $\qquad$ 21.05\\ 
              
\hline \hline 
\end{tabular}
\label{tab:experiments}
\end{center}
\end{table*}

\clearpage

\clearpage 

\clearpage 

\clearpage 

\clearpage 

\clearpage 

\clearpage

\clearpage

\clearpage 

\clearpage 

\end{document}